\documentclass[aps,prd,showpacs,twocolumn,superscriptaddress,showkeys,nofootinbib,
]{revtex4-1}

\usepackage{amsmath}
\usepackage{amsfonts}
\usepackage{amssymb}
\usepackage{bm}
\usepackage{dsfont}
\usepackage{epsfig}
\usepackage{color}
\usepackage{epstopdf}
\usepackage{hyperref}
\usepackage{lineno}

    \def\la{\lambda} \def\de{\delta}   \def\dag{\dagger}
\def\lap{\lambda^{\prime}} \def\pa{\partial} 
\def\bnabla{{\bm \nabla}}

\begin{document}

\preprint{TUM-EFT 95/17}

\title{Spin structure of heavy-quark hybrids}

\author{Nora Brambilla}
\affiliation{Physik-Department, Technische Universit\"at M\"unchen, 
James-Franck-Stra\ss e 1, 85748 Garching, Germany}
\affiliation{Institute for Advanced Study, Technische Universit\"at 
M\"unchen, Lichtenbergstrasse 2a, 85748 Garching, Germany}

\author{Wai Kin Lai}
\affiliation{Physik-Department, Technische Universit\"at M\"unchen, 
James-Franck-Stra\ss e 1, 85748 Garching, Germany}

\author{Jorge Segovia}
\affiliation{Grup de F\'\i sica Te\`orica, Dept. F\'\i sica and IFAE-BIST, 
Universitat Aut\`onoma de Barcelona, \\ E-08193 Bellaterra (Barcelona), Spain}

\author{Jaume Tarr\'us Castell\`a}
\affiliation{Grup de F\'\i sica Te\`orica, Dept. F\'\i sica and IFAE-BIST, 
Universitat Aut\`onoma de Barcelona, \\ E-08193 Bellaterra (Barcelona), Spain}

\author{Antonio Vairo}
\affiliation{Physik-Department, Technische Universit\"at M\"unchen, 
James-Franck-Stra\ss e 1, 85748 Garching, Germany}

\date{\today}

\begin{abstract}
  A unique feature of quantum chromodynamics (QCD), the theory of strong interactions, is the possibility for gluonic degrees of freedom to participate in the construction of physical hadrons, which are color singlets, in an analogous manner to valence quarks. Hadrons with no valence quarks are called glueballs, while hadrons where both gluons and valence quarks combine to form a color singlet are called hybrids. The unambiguous identification of such states among the experimental hadron spectrum has been thus far not possible. Glueballs are particularly difficult to establish experimentally since the lowest lying ones are expected to strongly mix with conventional mesons. On the other hand, hybrids should be easier to single out because the set of quantum numbers available to their lowest excitations may be exotic, i.e., not realized in conventional quark-antiquark systems. Particularly promising for discovery appear to be heavy hybrids, which are made of gluons and a heavy-quark-antiquark pair (charm or bottom). In the heavy-quark sector systematic tools can be used that are not available in the light-quark sector. In this paper we use a nonrelativistic effective field theory to uncover for the first time the full spin structure of heavy-quark hybrids up to $1/m^2$-terms in the heavy-quark-mass expansion. We show that such terms display novel characteristics at variance with our consolidated experience on the fine and hyperfine splittings in atomic, molecular and nuclear physics. We determine the nonperturbative contributions to the matching coefficients of the effective field theory by
comparing our results to lattice-QCD determinations of the charmonium hybrid spectrum and extrapolate the results to the bottomonium hybrid sector where lattice-QCD determinations are still challenging.
\end{abstract}

\maketitle

\noindent\emph{Introduction}.\,---\,Quantum chromodynamics (QCD), the strong-interaction part of the standard model of particle physics, presents a unique problem: the elementary degrees of freedom of the theory, quarks and gluons, are not the degrees of freedom accessible via the experiments. These elementary degrees of freedom are confined inside color-singlet states generically named hadrons, which form the spectrum of the theory. A very successful classification scheme for hadrons was independently proposed by Murray Gell-Mann~\cite{GellMann:1964nj} and George Zweig~\cite{Zweig:1964CERN} in 1964. It is called the quark model and classifies hadrons according to their (valence) quark content. Two main families were identified: mesons, made of a quark and an antiquark, and baryons, which contain three quarks. Nevertheless, QCD also allows for the existence of states formed by more than three valence quarks or by solely gluonic excitations (glueballs) or by gluonic excitations bound with quarks and antiquarks (hybrids).

The quark model received experimental verification in the late 1960s and, despite extensive searches, a hadron not fitting the quark-model classification scheme was not unambiguously identified until the last decade, the first being the $X(3872)$ discovered by the Belle Collaboration in 2003~\cite{Choi:2003ue}. Nowadays, more than two dozen nontraditional charmonium- and bottomoniumlike states, the so-called XYZ mesons, have been observed at experiments at B-factories (BaBar, Belle and CLEO), $\tau$-charm facilities (CLEO-c and BESIII) and also proton-(anti)proton colliders (CDF, D0, LHCb, ATLAS and CMS). There is growing evidence that at least some of the new states are nonconventional hadrons such as glueballs, hybrids or multiquark systems (see, e.g., reviews~\cite{Brambilla:2004wf,Brambilla:2010cs, Brambilla:2014jmp,Olsen:2014qna} for more details on the experimental and theoretical status of the subject). While multiquark states may, under some circumstances, have an electromagnetic analogue in molecular states, the discovery of a hadron containing a gluonic excitation would confirm one of the most unique features of QCD and ultimately open a new window on how matter is made.

In addition to the experiments mentioned above, many of which will remain active in the next two decades, a new experiment, PANDA@FAIR, has been designed to produce at the primary collision point a rich environment of gluons in order to promote the formation of glueballs and heavy hybrids~\cite{Lutz:2009ff}.\footnote{The experimental search of hybrid mesons in the light-quark sector has instead a long history mostly developed by the BNL-E852 and COMPASS collaborations and will continue with the new experiment GlueX@JLab12~\cite{Dudek:2012vr}.} 

To support the experimental effort different theoretical tools have been developed for studying this kind of bound state. In the early days of hadron phenomenology several models were proposed for glueballs and hybrids~\cite{Horn:1977rq, Barnes:1981ac, Chanowitz:1982qj, Barnes:1982tx, Cornwall:1982zn, Isgur:1984bm}, and more sophisticated versions have been developed in the last years~\cite{Kalashnikova:1993xb, Swanson:1998kx, Brau:2004xw, Szczepaniak:2005xi, Buisseret:2006wc, Guo:2008yz}. QCD sum rules have also been applied to the study of heavy hybrids~\cite{Govaerts:1984hc, Govaerts:1985fx, Govaerts:1986pp, Harnett:2012gs, Berg:2012gd, Chen:2013zia}. The interested reader is referred to~\cite{Lebed:2016hpi} and~\cite{Briceno:2015rlt} for reviews on hybrids.

Approaches fully rooted in QCD are lattice simulations and effective field theories (EFTs). The implementation of these two techniques is, however, not devoid of difficulties. For lattice calculations it is challenging to address excited states, which require a large operator basis leaving nevertheless sometimes ambiguities in the ultimate identification of the states. Studies of hybrids in lattice QCD have traditionally focused on the charmonium sector. Calculations in the bottomonium sector require smaller lattice spacings, making them computationally more demanding. A pioneering quenched calculation of the excited charmonium spectrum was presented in Ref.~\cite{Dudek:2007wv}. This study was extended by the RQCD collaboration~\cite{Bali:2011dc, Bali:2011rd} and by the Hadron Spectrum Collaboration~\cite{Liu:2012ze, Cheung:2016bym}. The common result in all these studies has been the identification of the lowest hybrid charmonium spin multiplet at about $4.3\,\text{GeV}$ containing a state with exotic quantum numbers $J^{PC}=1^{-+}$.

On the other hand, EFTs require clear separations between the relevant scales of the system in order to be useful, and necessitate the knowledge of matching coefficients often of nonperturbative nature. In QCD, $\Lambda_{\rm QCD}$ is the scale of a few hundred MeV at which nonperturbative effects dominate and a weak-coupling treatment is no longer valid. For heavy-quark or quarkonium hybrids, there is however at least one scale, the heavy-quark mass $m$, which is much larger than any other scale of the system and, in particular, larger than $\Lambda_{\rm QCD}$. The existence of this hierarchy of scales justifies the use of nonrelativistic effective field theories to factorize high-energy effects happening at the scale $m$ from low-energy ones. Moreover, the former may be computed in perturbation theory. Another hierarchy of scales is due to the existence of an energy gap between the gluonic excitations and the excitations of the heavy-quark-antiquark pair as observed in lattice data~\cite{Juge:1997nc, Bali:2000vr, Juge:2002br, Bali:2003jq}. The gluon dynamics is nonperturbative and, therefore, it occurs at the scale $\Lambda_{{\rm QCD}}$, while the nonrelativistic heavy-quark-antiquark pair binds in the background potential created by the gluonic excitations at an energy scale of order $mv^2$, where $v$ is the relative velocity between the heavy quark and antiquark~\cite{Brambilla:2004jw}. The observed energy gap requires the hierarchy of scales $\Lambda_{{\rm QCD}} \gg mv^2$ to be fulfilled. This energy gap, not present in light-quark hybrids, has led to the observation that quarkonium hybrids can be treated in the framework of the Born-Oppenheimer approximation~\cite{Griffiths:1983ah, Juge:1997nc,Juge:1999ie, Braaten:2014qka, Braaten:2014ita}. In recent papers, the Born-Oppenheimer approximation has been incorporated into an effective field theory formulation, which is both rigorous and systematic, and it has been used to compute the quarkonium hybrid spectrum~\cite{Berwein:2015vca, Oncala:2017hop, Brambilla:2017uyf}. The mixing of standard and hybrid quarkonia has also been studied in this framework in Ref.~\cite{Oncala:2017hop}.

A complete picture of the quarkonium hybrid spectrum cannot be obtained until, at least, the leading nonvanishing spin-dependent contributions are calculated. These terms break the degeneracy within the spin multiplets. The pattern and values of the spin splittings distinguishes different theoretical pictures for XYZ mesons; therefore their determination in the chamonium and bottomonium hybrid sectors will be an important step forward in the understanding of exotic quarkonium.

In this paper, combining the EFT suited for hybrids with information gained from lattice QCD, we derive for the first time the full set of spin-dependent hybrid potentials up to order $1/m^2$.\footnote{In Ref.~\cite{Soto:2017one}, relations between the spin splittings of different hybrid states were presented up to $1/m$-suppressed terms.} These include the well-known spin-orbit, spin-spin and tensor potentials that govern the fine and hyperfine splittings in a broad span of problems from atomic to nuclear and molecular physics, and a new set of interactions unique to heavy hybrids.

The spin-dependent potentials contain perturbative weakly coupled contributions and nonperturbative ones. The former can be computed analytically and are known up to some order in the strong coupling. The latter can be fitted on data or, in the absence of data, on the spectrum obtained from lattice QCD. Owing to the factorization of the EFTs, the whole heavy-quark flavor dependence of the spin-dependent potentials is encoded in the perturbative contributions, while nonperturbative contributions are flavor independent. 
Hence, we can use lattice data for charmonium hybrids to fit the nonperturbative contributions and predict the fine and hyperfine splittings of bottomonium hybrids, where precise lattice-QCD calculations have proved difficult.


\noindent\emph{Formalism}.\,---\,Following~\cite{Berwein:2015vca} we classify quarkonium hybrids according to their short-distance behavior   
in the static limit, i.e., at small heavy-quark-antiquark distances, $r \ll 1/\Lambda_{{\rm QCD}}$. There, quarkonium hybrids reduce to gluelumps, which are bound states made of a color-octet heavy-quark-antiquark pair and some gluonic fields in a color-octet configuration localized at the center of mass of the heavy-quark-antiquark pair~\cite{Foster:1998wu, Brambilla:1999xf, Berwein:2015vca}. A basis of gluelump states can be written as 
\begin{align}
|\kappa,\,\lambda\rangle =P^i_{\kappa\lambda} \, O^{a\,\dagger}\left(\bm{r},\bm{R}\right) \, G_{\kappa}^{ia}(\bm{R})|0\rangle\,,
\label{eigen1}
\end{align}
where $\bm{R}$ and $\bm{r}$ are the center-of-mass coordinate and the relative distance of the heavy-quark-antiquark pair, respectively; $P^i_{\kappa\lambda}$ is an operator that projects the gluelump to an eigenstate of $\bm{K}\cdot\hat{\bm{r}}$ with eigenvalue $\lambda$, where $\bm{K}$ is the angular momentum operator of the gluons and $\hat{\bm{r}}$ the unit vector along the heavy-quark-antiquark axis; $O^{a}$ is a color-octet heavy-quark-antiquark field and $G_{\kappa}^{ia}$ are gluonic fields in a color-octet configuration; $\kappa=K^{PC}$ with $P$ and $C$ parity and charge conjugation of the gluonic fields, and $i$ and $a$ are spin and color indices, respectively. Summations over the indices $i$ and $a$ are implied. The states $|\kappa,\,\lambda\rangle$ live in representations, characterized by $\lambda$, of the cylindrical-symmetry group $D_{\infty h}$ (with $P$ replaced by $CP$), which is the same symmetry group of diatomic molecules. The fields $\Psi_{\kappa\lambda}(t,\,\bm{r},\,\bm{R})$ associated to the states $|\kappa,\,\lambda\rangle$ are the natural degrees of freedom of the low-energy EFT for hybrids~\cite{Berwein:2015vca, Brambilla:2017uyf}. We use the same notation of~\cite{Brambilla:2017uyf} and refer to this EFT as the Born-Oppenheimer EFT (BOEFT).

The BOEFT describes the low-energy excitations ($E\sim mv^2$) of the hybrids. It is obtained by sequentially integrating out the modes at the higher energy scales $m$, $mv$ and $\Lambda_{{\rm QCD}}$. Here we extend the BOEFT of~\cite{Berwein:2015vca, Brambilla:2017uyf} to include spin-dependent terms of order $1/m$ and $1/m^2$. The Lagrangian describing the fields $\Psi_{\kappa\lambda}(t,\,\bm{r},\,\bm{R})$ reads
\begin{align}
L_{\rm BOEFT} \!&= \!\!\int \! d^3R\,d^3r  \sum_{\lambda\lambda^{\prime}}\Psi^{\dagger}_{\kappa\lambda}(t,\,\bm{r},\,\bm{R}) 
\biggl\{i\partial_t - V_{\kappa\lambda\lambda^{\prime}}(\bm{r}) \nonumber \\ 
\!& +P^{i\dag}_{\kappa\lambda}\frac{\bnabla^2_r}{m}P^i_{\kappa\lambda^{\prime}}\biggr\}\Psi_{\kappa\lambda^{\prime}}(t,\,\bm{r},\,\bm{R})+\dots\,,
\label{bolag2}
\end{align}
where the ellipsis stands for possible terms mixing hybrids with different quantum numbers $\kappa$, and hybrids with quarkonia. It also stands for hybrids coupling with light hadrons. The hybrids that we examine in this work are the lowest-lying excitations of $\kappa=1^{+-}$, which are separated by a gap of order $\Lambda_{\rm QCD}$ from excitations of different $\kappa$ and ordinary quarkonia. Hence, in our case, the contributions coming from other hybrid states and quarkonia are integrated out and included in the potential. They do not show up as explicit degrees of freedom in the Lagrangian \eqref{bolag2}, and the ellipsis stands only for hybrids coupling with light hadrons, which may be neglected. The potential $V_{\kappa\lambda\lambda^{\prime}}$ can be organized into an expansion in $1/m$ and a sum of spin-dependent ({\rm SD}) and spin-independent ({\rm SI}) parts:
\begin{eqnarray}
V_{\kappa\la\lap}(\bm{r}) &=& V^{(0)}_{\kappa\la}(r)\de_{\la\lap}+\frac{V^{(1)}_{\kappa\la\lap}(\bm{r})}{m}+\frac{V^{(2)}_{\kappa\la\lap}(\bm{r})}{m^2}+\dots, \quad \\
V_{\kappa\la\lap}^{(1)}(\bm{r}) &=& V_{\kappa\la\lap\,{\rm SD}}^{(1)}(\bm{r})+V_{{\kappa\la\lap}\,{\rm SI}}^{(1)}(\bm{r})\,,\\
V_{\kappa\la\lap}^{(2)}(\bm{r}) &=& V_{\kappa\la\lap\,{\rm SD}}^{(2)}(\bm{r})+V_{{\kappa\la\lap}\,{\rm SI}}^{(2)}(\bm{r})\,.
\end{eqnarray}

Besides the static potential, $V^{(0)}_{\kappa\la}(r)$, the other potentials have not been computed in lattice QCD. We gain some information by looking at them at short distances: $r \Lambda_{\rm QCD} \ll 1$. At short distances the potentials may be organized as a sum of a perturbative part, which is typically nonanalytic in $r$, and a nonperturbative part, which is a series in powers of $r$. The perturbative part comes from integrating out modes scaling with the inverse quark-antiquark distance $1/r$, which at short distances is much larger than $\Lambda_{{\rm QCD}}$. The nonperturbative part comes from integrating out modes scaling with $\Lambda_{{\rm QCD}}$. The short-distance expansion is consistent with our quantum number attribution to the states. It is limited, however, to the lowest-lying, most compact, hybrid states, the ones most likely to be described in terms of gluelumps.

For the lowest-lying hybrid excitations of $\kappa=1^{+-}$,\footnote{According to lattice QCD the hybrid static potentials $V^{(0)}_{1^{+-}\la}(r)$ are the lowest lying at short distances~\cite{Juge:1997nc,Juge:2002br}.} the spin-dependent potentials take the form 
\begin{eqnarray}
V_{1^{+-}\la\lap\,{\rm SD}}^{(1)}(\bm{r}) &=& V_{{\rm SK}}(r)\left(P^{i\dag}_{1\la}\bm{K}^{ij}P^j_{1\lap}\right)\cdot\bm{S}  \nonumber\\
&& \hspace{-1cm} + V_{{\rm SK}b}(r)\left[\left(\bm{r}\cdot \bm{P}^{\dag}_{1\la}\right)\left(r^i\bm{K}^{ij}P^j_{1\lap}\right)\cdot\bm{S}\right. \nonumber \\
&& \hspace{-1cm} -\left.\left(r^i\bm{K}^{ij}P^{j\dag }_{1\la}\right)\cdot\bm{S} \left(\bm{r}\cdot \bm{P}_{1\lap}\right)\right]+\dots, \label{sdm2}\\
V_{1^{+-}\la\lap\,{\rm SD}}^{(2)}(\bm{r}) &=& V_{{\rm SL}a}(r)\left(P^{i\dag}_{1\la}\bm{L}_{Q\bar{Q}}P^i_{1\lap}\right)\cdot\bm{S}\nonumber\\
&& \hspace{-1cm} + V_{{\rm SL}b}(r)P^{i\dag}_{1\la}\left(L_{Q\bar{Q}}^iS^j+S^iL_{Q\bar{Q}}^j\right)P^{j}_{1\lap}\nonumber\\
&& \hspace{-1cm} + V_{{\rm SL}c}(r)\left[\left(\bm{r}\cdot \bm{P}^{\dag}_{1\la}\right)\left(\bm{p}\times\bm{S}\right)\cdot \bm{P}_{1\lap}\right.\nonumber\\
&& \hspace{-1cm} + \left.\bm{P}^{\dag }_{1\la}\cdot\left(\bm{p}\times\bm{S}\right)\left(\bm{r}\cdot \bm{P}_{1\lap}\right)\right]\nonumber\\
&& \hspace{-1cm} + V_{{\rm S}^2}(r)\bm{S}^2\de_{\la\lap}+V_{{\rm S}_{12}a}(r)S_{12}\de_{\la\lap}\nonumber\\
&& \hspace{-1cm} + V_{{\rm S}_{12}b}(r)P^{i\dag}_{1\la}P^j_{1\lap}\left(S^i_1S^j_2+S^i_2S^j_1\right)+\dots,\quad\label{sdm3}
\end{eqnarray}
where $\bm{L}_{Q\bar{Q}}$ is the orbital angular momentum of the heavy-quark-antiquark pair, $\bm{S}_1$ and $\bm{S}_2$ are the spin vectors of the heavy quark and heavy antiquark, respectively, $\bm{S}=\bm{S}_1+\bm{S}_2$ and ${S}_{12}=12(\bm{S}_1\cdot\hat{\bm{r}})(\bm{S}_2\cdot\hat{\bm{r}})-4\bm{S}_1\cdot\bm{S}_2$. $\left({K}^{ij}\right)^k=i\epsilon^{ikj}$ is the angular momentum operator for the spin-1 gluons. The projectors $P^{i}_{1\lambda}$ read
\begin{align}
P^{i}_{10}&=\hat{r}^i_0=\hat{r}^i \,,\label{pr10}\\
P^{i}_{1\pm 1}&=\hat{r}^i_\pm=\mp\left(\hat{\theta}^i\pm i\hat{\phi}^i\right)/\sqrt{2}\,,\label{pr11}
\end{align}
where $\hat{\bm r} = (\sin\theta\cos\phi,\,\sin\theta\sin\phi\,,\cos\theta)$, $\hat{\bm \theta} =$ $(\cos\theta\cos\phi,$ $\,\cos\theta\sin\phi\,,-\sin\theta)$ 
and $\hat{\bm \phi} = (-\sin\phi,\,\cos\phi\,,0)$. The ellipses in Eqs. \eqref{sdm2} and \eqref{sdm3} stand for terms suppressed by powers of $r\Lambda_{\rm QCD}$.

The matching coefficients of the spin-dependent potentials read
\begin{eqnarray}
V_{{\rm SK}}  &=& V^{\rm np}_{{\rm SK}}\,,\label{vsk}\\
V_{{\rm SK}b} &=& V^{\rm np}_{{\rm SK}b}\,,\label{vskb}\\
V_{{\rm SL}a} &=& V_{o\,{\rm SL}} + V^{\rm np}_{{\rm SL}a}\,,\label{vsla}\\
V_{{\rm SL}b} &=& V^{\rm np}_{{\rm SL}b}\,,\label{vslb}\\
V_{{\rm SL}c} &=& V^{\rm np}_{{\rm SL}c}\,,\label{vps}\\
V_{{\rm S}^2} &=& V_{o\,{\rm S}^2} + V^{\rm np}_{{\rm S}^2}\,, \label{vs2}\\
V_{{\rm S}_{12}a} &=& V_{o\,{\rm S}_{12}}\,, \label{vs12a}\\
V_{{\rm S}_{12}b} &=& V^{\rm np}_{{\rm S}_{12}b}\,,\label{vs12b}
\end{eqnarray}
where we have separated the perturbative parts, $V_{o\,{\rm SL}}$, $V_{o\,{\rm S}^2}$ and $V_{o\,{\rm S}_{12}}$, from the nonperturbative ones, labeled with ``np''. The perturbative parts are the spin-dependent parts of the color-octet quark-antiquark potential, computed at leading order in~\cite{Pineda:1996nw}. The nonperturbative parts can be written as an expansion in $r^2$: $V^{\rm np}=V^{\rm np\,{\rm(0)}}+V^{\rm np\,{\rm (1)}}r^2+\dots$. The coefficients $V^{\rm np\,{\rm (i)}}$ do not depend on $r$ and can be expressed in terms of gluonic correlators. They are made out of the gluon fields in the gluelump and in the interaction vertices of the weakly coupled potential nonrelativistic QCD (pNRQCD) Lagrangian that can be read off from Ref.~\cite{Brambilla:2003nt}. Eventually, gluonic correlators may be computed on the lattice. For the purpose of this paper we consider them just as constants independent of each other that respect, however, the power counting of the EFT.

We aim to include terms up to order $\Lambda_{{\rm QCD}}^3/m^2$ and $mv^4$ to the spin splittings. The perturbative spin-dependent potentials $V_{o\,{\rm SL}}$, $V_{o\,{\rm S}^2}$ and $V_{o\,{\rm S}_{12}}$ scale like $m^3v^4$ if we count the Coulomb potential as $mv^2$. For the nonperturbative potential  $V^{\rm np}_{\rm SK}$ we need to include $V^{\rm np\,{\rm (0)}}_{\rm SK}$, which scales like  $\Lambda^2_{{\rm QCD}}$, and  $V^{\rm np\,{\rm (1)}}_{\rm SK}$, which scales like $\Lambda^4_{{\rm QCD}}$. Higher order terms in the expansion of $V^{\rm np}_{\rm SK}$ may or may not be relevant in dependence of the relative size of $\Lambda_{{\rm QCD}}/m$ with respect to $v$. In the present analysis we neglect them.
 The size of the other potentials follows from dimensional analysis: $V^{\rm np\,(0)}_{{\rm SL}a},V^{\rm np\,(0)}_{{\rm SL}b},
V^{\rm np\, (0)}_{{\rm SL}c},V^{\rm np\,(0)}_{{\rm S}^2} ,V^{\rm np\,(0)}_{{\rm S}_{12}b}\sim \Lambda^3_{{\rm QCD}}$, and $V^{\rm np\,{\rm (0)}}_{{\rm SK}b}\sim \Lambda^4_{{\rm QCD}}$. Spin-independent potentials are irrelevant at our accuracy. Only the $r$-dependent part of the spin-independent $1/m$ potential could, in principle, contribute to the spin splittings in second order perturbation theory if it is of order $\Lambda_{\rm QCD}^2/m$ or larger. However, the nonperturbative contribution is necessarily $r^2 \Lambda_{\rm QCD}^4/m \sim \Lambda_{\rm QCD}^2/m \times (\Lambda_{\rm QCD} r)^2 \ll \Lambda_{\rm QCD}^2/m$ and the perturbative one shows up only at one loop and is therefore of order $m v^4$ as in the quarkonium case~\cite{Brambilla:2000gk}.

\begin{figure}[ht]
\centerline{\includegraphics[width=.25\textwidth]{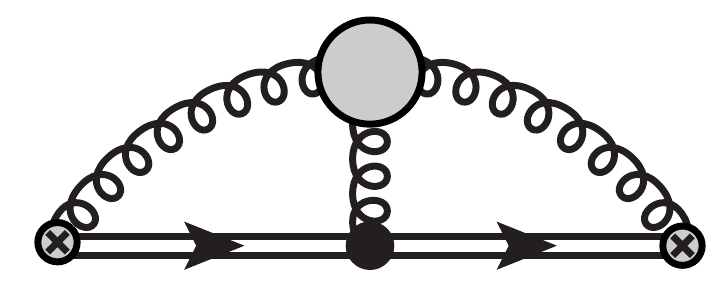}}
\caption{Feynman diagram contributing to $V^{\rm np\,({\rm 0})}_{SK}$. The double and curly lines represent the heavy-quark-antiquark octet and the gluon field respectively. The black dot stands for a $c_F \bm{S}_i\cdot \bm{B}/m$ induced vertex from the weakly coupled pNRQCD Lagrangian, the circles with a cross for the gluelump operators and the shaded circle represents nonperturbative gluon exchanges.}
\label{md}
\end{figure}

The flavor dependence of the BOEFT Lagrangian is in the mass $m$ and, at the order we are working here, in the one-loop expression of the quark chromomagnetic matching coefficient $c_F$ that enters $V^{\rm np\,({\rm 0})}_{SK}$; see the Feynman diagram in Fig.~\ref{md}. The expression of $c_F$ at $\mathcal{O}(\alpha_{\rm s})$ has been derived in Ref.~\cite{Eichten:1990vp}. 

In Ref.~\cite{Berwein:2015vca} the coupled Schr\"odinger equations resulting from the Lagrangian~\eqref{bolag2} with the mixing of the static potentials $V^{(0)}_{1^{+-}\la}(r)$, $\lambda=0,\,\pm1$, have been solved and the hybrid spectrum has been calculated. The mixing with other static states is suppressed by a large energy gap. The static potentials $V^{(0)}_{1^{+-}\la}(r)$ have been obtained from matching it to the static energies computed on the lattice in Refs.~\cite{Bali:2003jq,Juge:2002br}. There are two types of solution corresponding to states with opposite parity, $\Psi^{N j m_j l s }_{+}$ and $\Psi^{N j m_j l s }_{-}$,\footnote{The $P$ and $C$ of $\Psi_{\pm}$ are $P=\pm(-1)^l$ and $C=\pm(-1)^{l+s}$.}
\begin{eqnarray}
&& \Psi^{N j m_j l s }_{+}(\bm{r}) \nonumber\\
&=&\sum_{m_l m_s} \mathcal{C}^{j m_j}_{l\,m_l\,s\,m_s}\left(
\begin{array}{c}
\psi_0^{(N)}(r)v_{l\,m_l}^0(\theta,\phi) \\
\frac{1}{\sqrt{2}}\psi_{+}^{(N)}(r)v_{l\,m_l}^{+1}(\theta,\phi) \\
\frac{1}{\sqrt{2}}\psi_{+}^{(N)}(r)v^{-1}_{l\,m_l}(\theta,\phi) \\
\end{array}
\right)\chi_{s\,m_s},\;\quad\label{psip} \\
&& \Psi^{N j m_j l s }_{-}(\bm{r}) \nonumber\\
&=&\sum_{m_l m_s} \mathcal{C}^{j m_j}_{l\,m_l\,s\,m_s}\left(
\begin{array}{c}
0 \\
\frac{1}{\sqrt{2}}\psi_{-}^{(N)}(r)v_{l\,m_l}^{+1}(\theta,\phi) \\
-\frac{1}{\sqrt{2}}\psi_{-}^{(N)}(r)v_{l\,m_l}^{-1}(\theta,\phi) \\
\end{array}
\right)\chi_{s\,m_s},\;\quad\label{psim}
\end{eqnarray}
where $l(l+1)$ is the eigenvalue of $\bm{L}^2$, with $\bm{L}=(\bm{L}_{Q\bar{Q}}+\bm{K})$; $j(j+1)$ and $m_j$ are the eigenvalues of $\bm{J}^2$ and $J_3$ respectively, with $\bm{J}=\bm{L}+\bm{S}$; $s(s+1)$ is the eigenvalue of $\bm{S}^2$; the $\mathcal{C}^{j m_j}_{l\,m_l\,s\,m_s}$ are Clebsch-Gordan coefficients. The eigenfunctions $v_{l\,m_l}^{\lambda}$ are generalizations of the associated Legendre polynomials for systems with cylindrical symmetry. Their derivation can be found in textbooks such as Ref.~\cite{LandauLifshitz}. The $\chi_{s\,m_s}$ are the spin wave functions. The radial wave functions $\psi_{0}^{(N)},\,\psi_{+}^{(N)},\,\psi_{-}^{(N)}$ are obtained numerically by solving the coupled Schr\"odinger equations, with $N$ labeling the radially excited states.


\noindent\emph{Results}.\,---\,We use standard time-independent perturbation theory to compute the mass shifts in the heavy hybrid spectrum produced by the spin-dependent operators in Eqs.~\eqref{sdm2} and \eqref{sdm3} for the states with wave functions given in Eqs.~\eqref{psip} and \eqref{psim}. We carry out perturbation theory to second order for the $1/m$-suppressed operator proportional to $V_{\rm SK}^{\rm np\,(0)}$ in Eq.~\eqref{sdm2}, and to first order for the other operators in Eqs.~\eqref{sdm2} and \eqref{sdm3}. Since the orbital wave functions $v^{\lambda}_{l m_l}(\theta,\,\phi)$ are eigenfunctions of $\bm{L}^2=\left(\bm{L}_{Q\bar{Q}}+\bm{K}\right)^2$ instead of $\bm{L}_{Q\bar Q}^2$, the computation of the expectation values of operators containing $\bm{L}_{Q\bar{Q}}$ requires some care. The details on the computation of these matrix elements can be found in Appendix~\ref{appendix}.

Finally, we note that the expectation values of $V_{o{\rm\,S^2}}(r)$ vanish for all our hybrid states. This follows from $V_{o\,{\rm S^2}}(r)\sim\de^3(\bm{r})$ and from the hybrid wave functions being 0 at the origin owing to the spin of the gluons in the gluelumps~\cite{Berwein:2015vca}.

\begin{table}[ht]
\caption{Lowest-lying quarkonium hybrid multiplets. The number labeling $H$ reflects the order in which the state appears in the spectrum from lower to higher masses. Note that the $l=0$ state is not the lowest mass state~\cite{Berwein:2015vca}.}
\begin{center}
\begin{tabular}{c|c|c|c}
\hline
\hline
Multiplet & $\,\,\,l\,\,\,$ & $J^{PC}(s=0)$ & $J^{PC}(s=1)$\\
\hline
$H_1$& $1$ & $1^{--}$ & $(0,1,2)^{-+}$ \\
$H_2$& $1$ & $1^{++}$ & $(0,1,2)^{+-}$ \\
$H_3$& $0$ & $0^{++}$ & $1^{+-}$ \\
$H_4$& $2$ & $2^{++}$ & $(1,2,3)^{+-}$ \\
\hline
\hline
\end{tabular}
\label{tb:spin_multiplet}
\end{center}
\end{table}

We present results for the four lowest-lying spin multiplets shown in Table~\ref{tb:spin_multiplet}. 
The eight nonperturbative parameters $V^{\rm np\,({\rm 0})}_{\rm SK}$, $V^{\rm np\,({\rm 1})}_{\rm SK}$, $V^{\rm np\,({\rm 0})}_{{\rm SK}b}$, 
$V^{\rm np\,(0)}_{{\rm SL}a}$, $V^{\rm np\,(0)}_{{\rm SL}b}$, $V^{\rm np\,(0)}_{{\rm SL}c}$, $V^{\rm np\,(0)}_{{\rm S}^2}$, $V^{\rm np\,(0)}_{{\rm S_{12}}b}$
that appear in the spin-dependent potentials in Eqs.~\eqref{vsk}-\eqref{vs12b} are obtained 
through a fitting procedure from comparing
the spin splittings resulting from our calculation
with the lattice data of the charmonium hybrid spectrum. 

Two sets of lattice data from the Hadron Spectrum Collaboration are available, one set from Ref.~\cite{Liu:2012ze} with a pion mass of $m_{\pi}\approx 400$~MeV and a more recent set from Ref.~\cite{Cheung:2016bym} with a pion mass of $m_{\pi}\approx 240$ MeV. Uncertainties associated to discretization effects were estimated in Ref.~\cite{Liu:2012ze} to be $\sim 40$~MeV. No analogous study was carried out in Ref.~\cite{Cheung:2016bym}, but similar associated uncertainties are expected. Adopting the same theoretical setting as in~\cite{Berwein:2015vca}, we define the heavy-quark masses in the RS$^\prime$ scheme~\cite{Pineda:2001zq}, which has the advantage to not affect the quark chromomagnetic matching coefficient, $c_F$, at the order we are working. For the charm mass we take $m^{\rm RS}_c(1{\rm GeV})=1.477$~GeV and for $\alpha_{\rm s}$ at four loops with three massless flavors, $\alpha_{\rm s}(2.6\textrm{~GeV})=0.26$.
In our
fitting procedure
the lattice data are weighted by $(\Delta^2_{\textrm{lattice}}+\Delta^2_{\textrm{high-order}})^{1/2}$, where $\Delta_{\textrm{lattice}}$ is
the uncertainty of the lattice data and $\Delta_{\textrm{high-order}}=(m_{\textrm{lattice}}-m_{\textrm{lattice\,spin-average}})\times\Lambda_{\rm QCD}/m$
is the estimated size of the theoretical uncertainty due to higher order terms in the potential, with $\Lambda_{\rm QCD}$ taken to be $0.5$~GeV. 
The size of the $V^{\rm np\,(i)}$'s is introduced
in the fitting procedure through a prior. 
The outcome
is consistent with the nonperturbative potentials scaling naturally, i.e., $V^{\rm np\,(i)} \sim ($few hundreds MeV$)^d$, where $d$ is the mass dimension of $V^{\rm np\,(i)}$.
\begin{figure*}[ht]
\begin{center}
\includegraphics[width=0.47\textwidth]{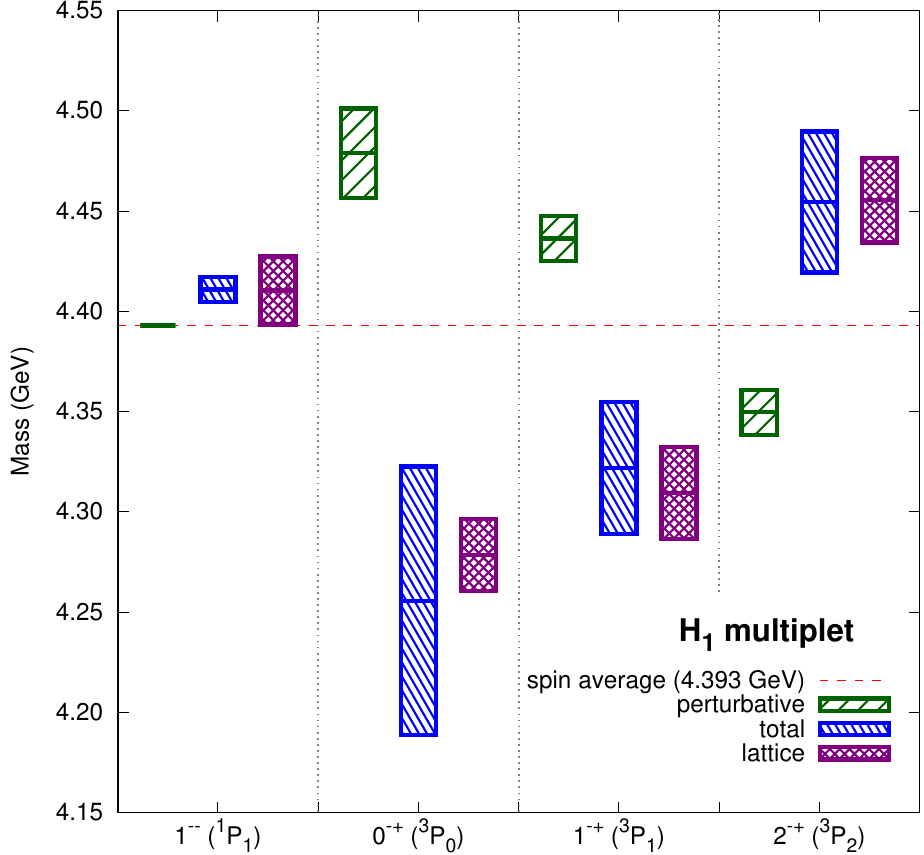}
\hspace*{0.50cm}
\includegraphics[width=0.47\textwidth]{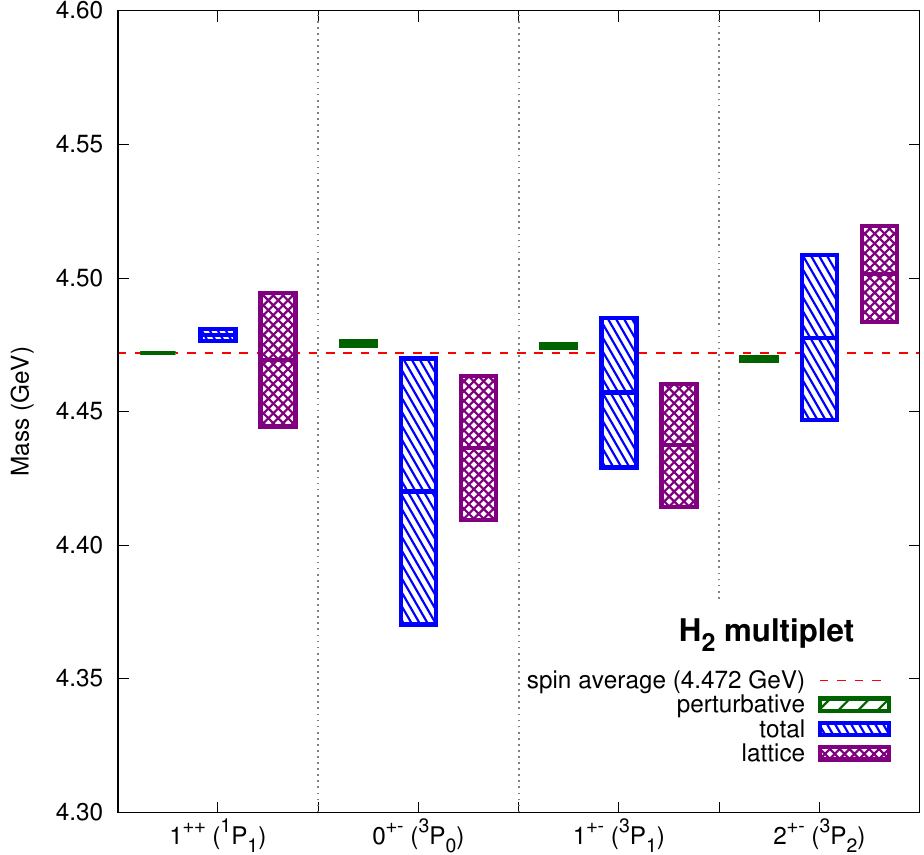} 
\\[2ex]
\includegraphics[width=0.47\textwidth]{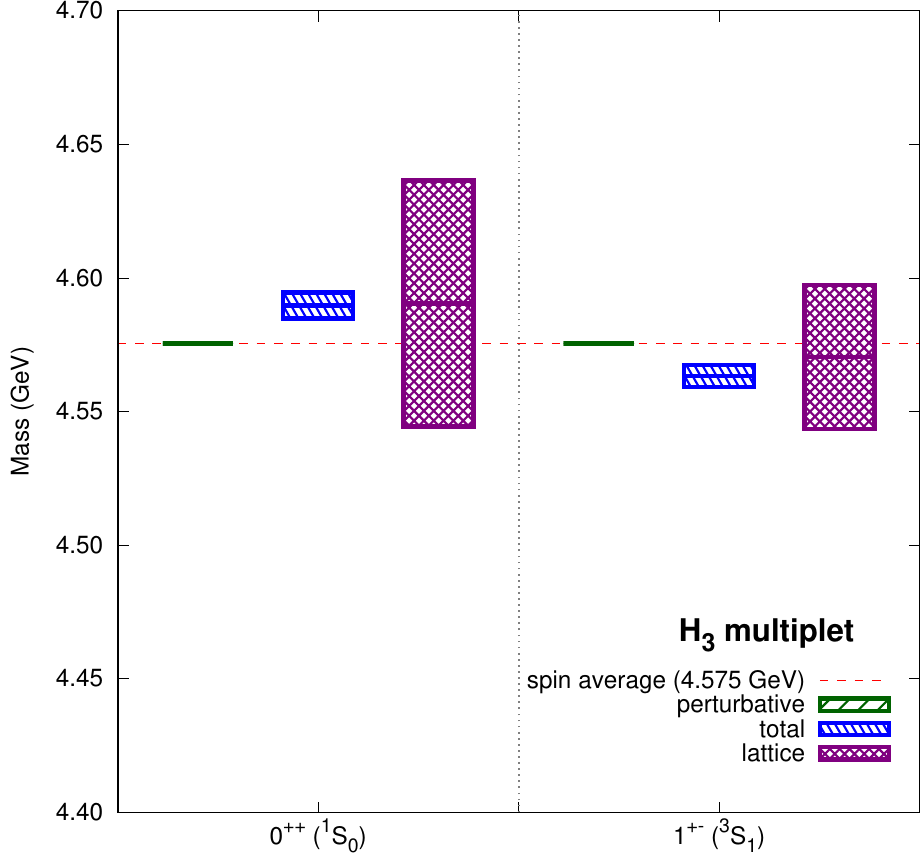}
\hspace*{0.50cm}
\includegraphics[width=0.47\textwidth]{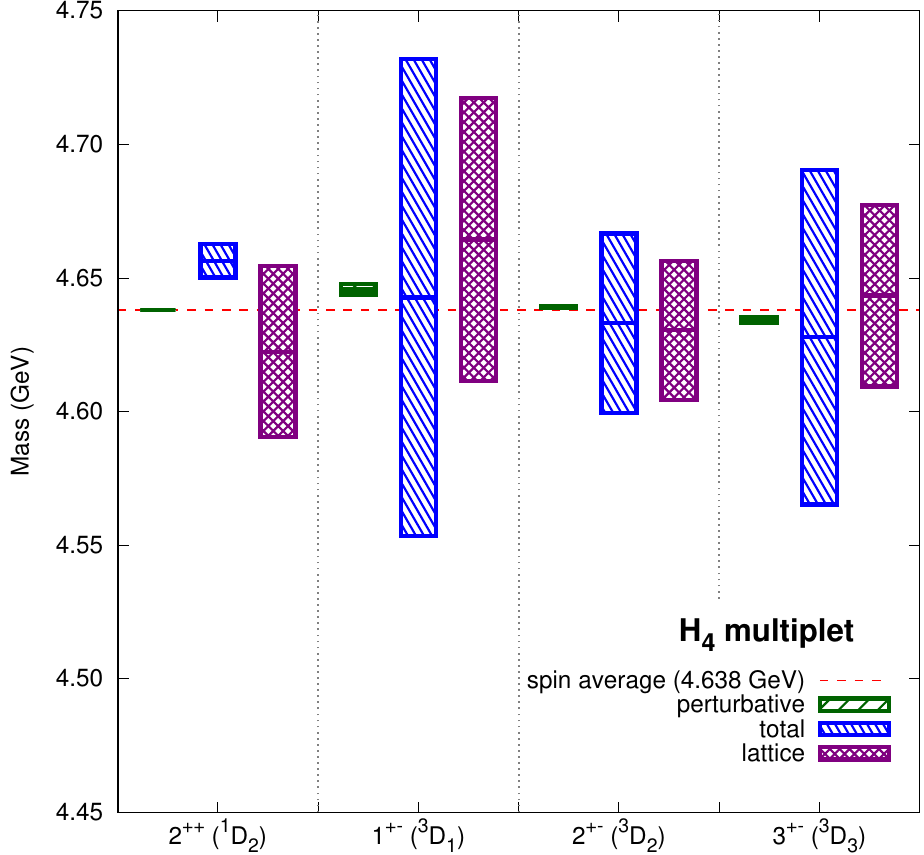}
\caption{Spectrum of the four lowest-lying charmonium hybrid multiplets. The lattice results from Ref.~\cite{Cheung:2016bym} with $m_{\pi}\approx 240$~MeV are the most right (purple) boxes for each quantum number. The perturbative contributions to the spin-dependent operators in Eq.~\eqref{sdm3} added to the spin average of the lattice results (red dashed lines) are the most left (green) boxes. The central (blue) boxes for each quantum number are the full results from the spin-dependent operators of Eqs.~\eqref{sdm2} and \eqref{sdm3} including perturbative and nonperturbative contributions. 
  The unknown nonperturbative matching coefficients are
  determined through a fitting procedure that reproduces
the lattice data. The height of the boxes indicates the uncertainty as detailed in the text.}
\label{fg:ccg_Cheung}
\end{center}
\end{figure*}

The results of
our fitting procedure applied
to the lattice data of Ref.~\cite{Cheung:2016bym} are shown in Fig.~\ref{fg:ccg_Cheung}.\footnote{
The resulting $\chi^2$/d.o.f. for the eight $V^{\rm np\,(i)}$'s is $0.99998$. Note that the leading contribution $V^{\rm np\,(0)}_{\rm SK}$ has the most dominant effect.}
The values of the nonperturbative matching coefficients obtained in
this way
are shown in Table~\ref{tb:npfit}. Each panel in Fig.~\ref{fg:ccg_Cheung} corresponds to one of the multiplets of Table~\ref{tb:spin_multiplet}. The most right (purple) boxes for each quantum number indicate the lattice results: the middle line corresponds to the mass of the state obtained from the lattice and the height of the box provides the uncertainty. The (red) dashed line indicates the spin-average mass of the lattice results. The most left (green) boxes for each quantum number correspond to the spin splittings from the perturbative contributions to Eqs.~\eqref{vsk}-\eqref{vs12b}. The height of these boxes ($\Delta_{\rm p}$) is an estimate of the uncertainty, given by the parametric size of the higher order corrections to the perturbative part of the potential, which is $\mathcal{O}(m\alpha_{\rm s}^5)$. The central (blue) boxes for each quantum number are the full results including the nonperturbative contributions after 
determining them through our fitting procedure.
The height of these boxes provides the uncertainty of the full result. This uncertainty is given by $\Delta_{\rm full}=(\Delta_{\rm p}^2 +\Delta^2_{\rm np}+\Delta^2_{\rm fit})^{1/2}$, where the uncertainty of the nonperturbative contribution $\Delta_{\rm np}$ is estimated to be of the same parametric size as the next order contribution to the matching coefficients. $\Delta_{\rm fit}$ is the statistical error of the fitting procedure.

The light-quark mass dependence of the spin splittings in hybrid charmonium can be studied comparing the results of Refs.~\cite{Liu:2012ze} and~\cite{Cheung:2016bym}, where only a mild light-quark mass dependence is observed and the pattern and number of charmonium states at the two pion masses are the same. Moreover the leading light-quark mass dependence of the hybrid spectrum cancels out in the case of the spin splittings. The light-quark dependence of the nonperturbative matching coefficients of our EFT can be estimated by
applying the same fitting procedure used for the data of Ref.~\cite{Cheung:2016bym} to
the lattice data of Ref.~\cite{Liu:2012ze} and comparing with the results presented in Fig.~\ref{fg:ccg_Cheung}. The precision of the determination of the nonperturbative parameters is limited to the size of the neglected higher order contribution in our EFT counting; we observe that the variation of the parameters is within or close to this theoretical uncertainty; therefore no significant light-quark dependence of the nonperturbative parameters can be inferred.

\begin{table}[!t]
\caption{Nonperturbative matching coefficients determined by comparing the charmonium hybrid spectrum obtained from the hybrid EFT to the lattice spectrum from the Hadron Spectrum Collaboration data of Ref.~\cite{Cheung:2016bym} with pion mass of $m_{\pi}\approx 240\textrm{ MeV}$, respectively. The matching coefficients are normalized to their parametric natural size. We take the value $\Lambda_{QCD}=0.5$ GeV.}
\begin{center}
\begin{tabular}{c|c}\hline\hline
$V^{{\rm np}\,(0)}_{\rm SK}/\Lambda^2_{QCD}$ & $1.03$\\
$V^{{\rm np}\,(1)}_{\rm SK}/\Lambda^4_{QCD}$ & $-0.51$\\
$V^{{\rm np}\,(0)}_{{\rm SK}b}/\Lambda^4_{QCD}$ & $0.28$\\
$V^{{\rm np}\,(0)}_{{\rm SL}a}/\Lambda^3_{QCD}$     & $-1.32$\\
$V^{{\rm np}\,(0)}_{{\rm SL}b}/\Lambda^3_{QCD}$     & $2.44$\\
$V^{{\rm np}\,(0)}_{{\rm SL}c}/\Lambda^3_{QCD}$     & $0.87$\\
$V^{{\rm np}\,(0)}_{{\rm S}^2}/\Lambda^3_{QCD}$     & $-0.33$\\
$V^{{\rm np}\,(0)}_{{\rm S_{12}}b}/\Lambda^3_{QCD}$ & $-0.39$\\
\hline\hline
\end{tabular}
\label{tb:npfit}
\end{center}
\end{table}

The multiplets $H_1$ and $H_2$ correspond to states with $l=1$, and negative and positive parity, respectively. The mass splitting between $H_1$ and $H_2$ is a result of the mixing between states of $\la=0$ and $\la=\pm1$ induced by the kinetic-energy operator~\cite{Berwein:2015vca}, a phenomenon known in molecular physics as $\Lambda$-doubling. The perturbative parts of the matching coefficients produce a pattern of spin splittings which is opposite to the one predicted by lattice-QCD simulations and opposite to ordinary quarkonia due to the repulsive nature of the octet potential. This discrepancy can be reconciled thanks to the nonperturbative contributions, in particular, due to the dominant one: $V_{\rm SK}^{\rm np\,(0)}$, which is of order $\Lambda_{\rm QCD}^2/m$, and, therefore, parametrically larger than the perturbative contributions, which are of order $mv^4$. A consequence of the countervail of the perturbative contributions is a relatively large uncertainty on the full result caused by the large nonperturbative contributions. Due to this uncertainty the mass hierarchies among the spin triplet states of some multiplets, in particular, for the $H_4$ charmonium hybrid multiplet, could not be firmly established.

\begin{figure*}[!t]
\begin{center}
\includegraphics[width=0.47\textwidth]{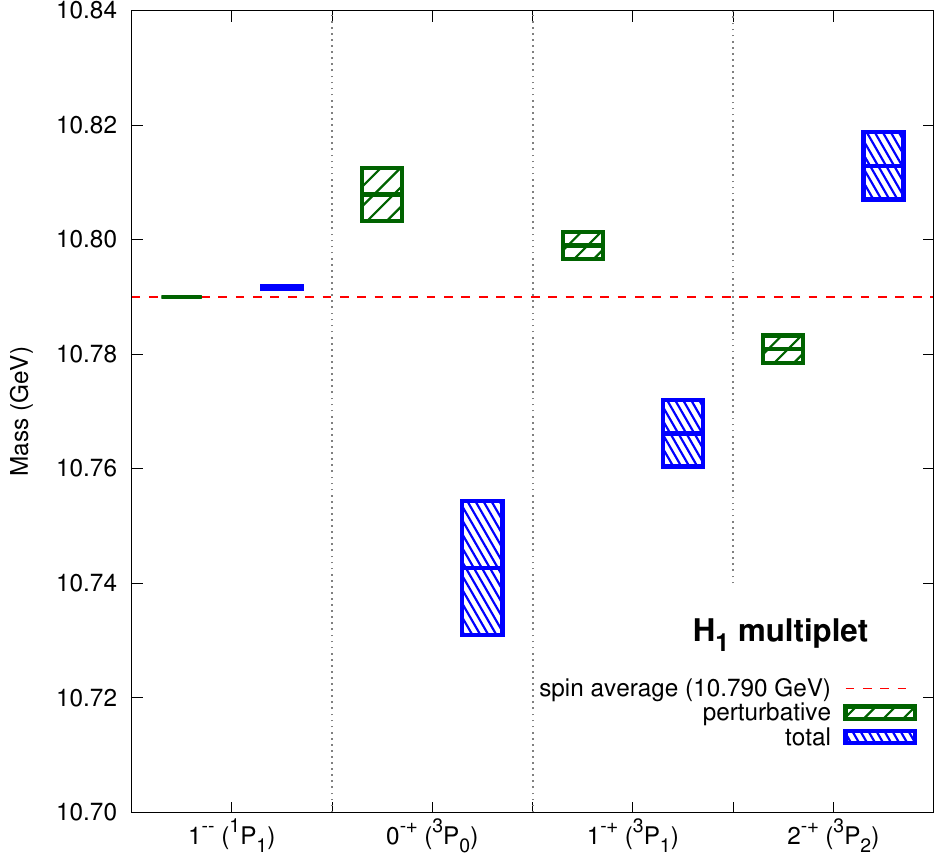}
\hspace*{0.50cm}
\includegraphics[width=0.47\textwidth]{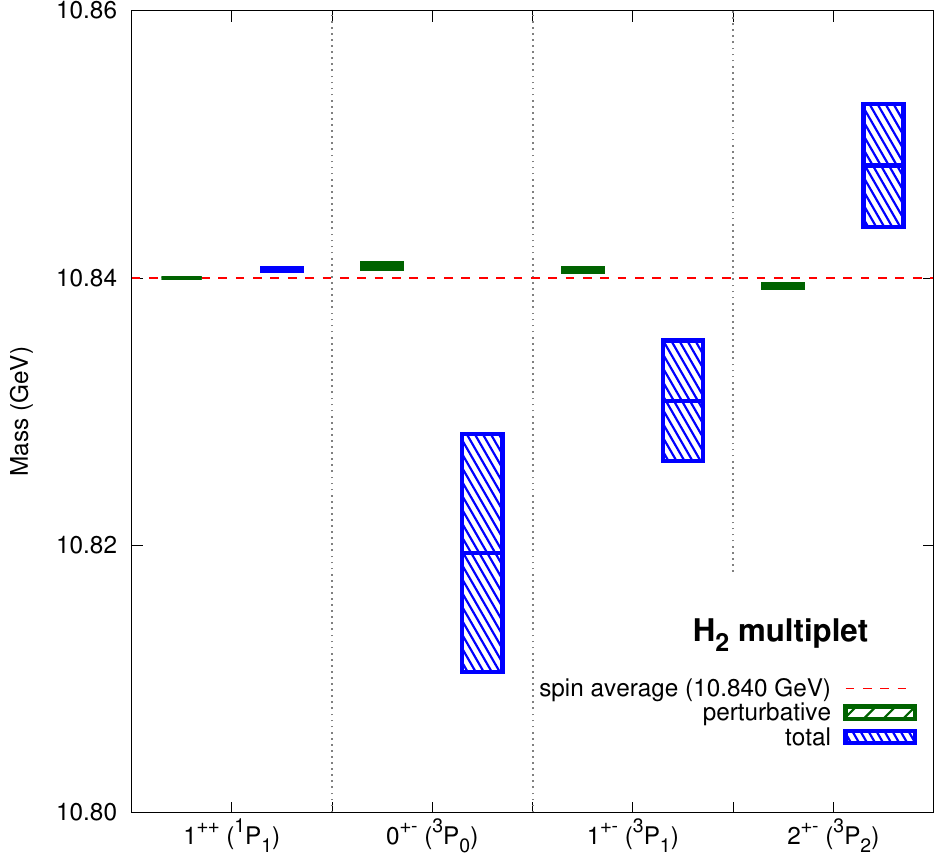} 
\\[2ex]
\includegraphics[width=0.47\textwidth]{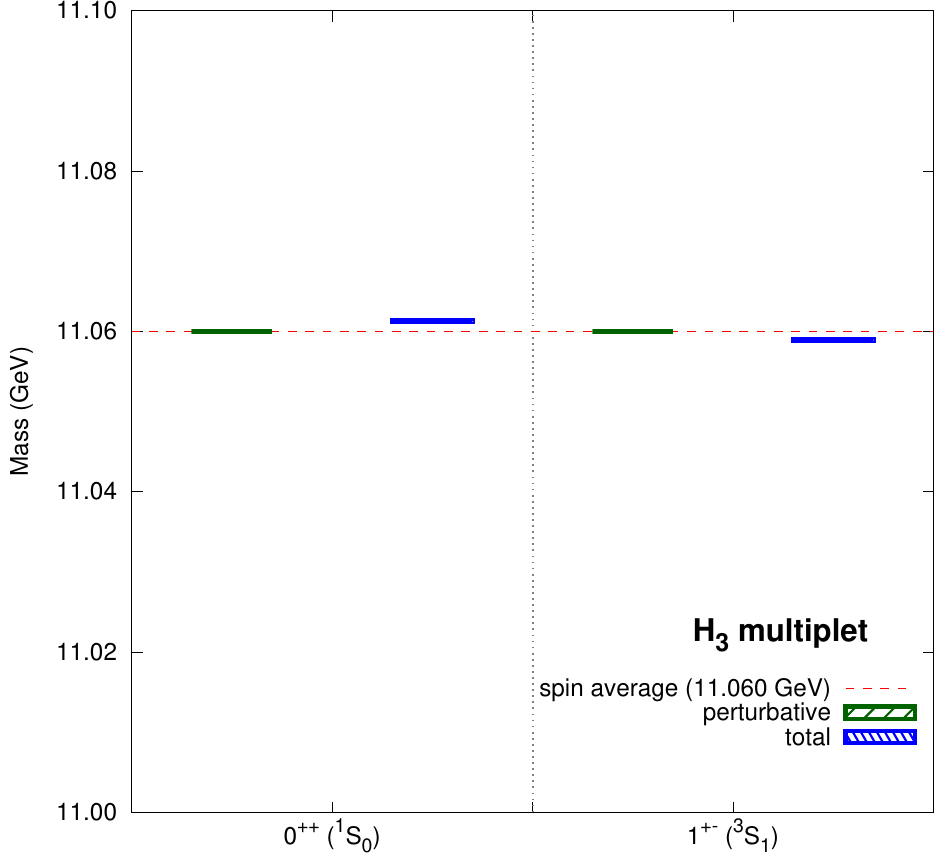}
\hspace*{0.50cm}
\includegraphics[width=0.47\textwidth]{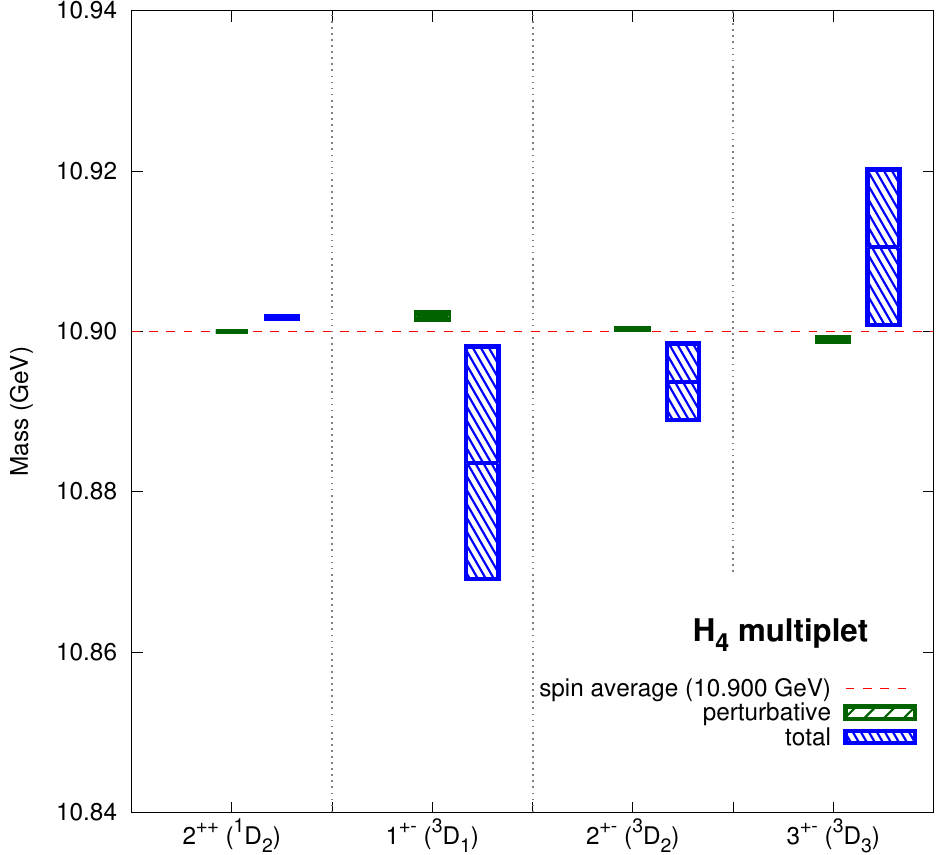}
\caption{Spectrum of the four lowest-lying bottomonium hybrids computed by adding the spin-dependent contributions from Eqs.~\eqref{vsk}-\eqref{vs12b} to the spectrum obtained in Ref.~\cite{Berwein:2015vca}. The nonperturbative contribution to the matching coefficients is determined from
    the charmonium hybrids spectrum of Ref.~\cite{Cheung:2016bym} shown in Fig.~\ref{fg:ccg_Cheung}. The average mass for each multiplet is shown as a red line. The results with only the perturbative contributions included and the full results with perturbative and nonperturbative contributions included are shown for each quantum number by the left (green) and the right (blue) boxes, respectively. The height of the boxes indicates the uncertainty as detailed in the text.}
\label{fg:bbg_Cheung}
\end{center}
\end{figure*}

Having determined the values of the $V^{\rm np\,(i)}$'s
from the charmonium hybrid spectrum computed on the lattice, we can predict the spin contributions in the bottomonium hybrid sector, for which lattice determinations are yet very sparse. The only flavor dependence of the $V^{\rm np\,(i)}$'s relevant at the precision we aim for is in the mass and in the quark chromomagnetic matching coefficient, $c_F$, in $V^{\rm np\,({\rm 0})}_{SK}$. Both are known. We compute the bottomonium hybrid spectrum by adding the spin-dependent contributions from Eqs.~\eqref{vsk}-\eqref{vs12b} to the spectrum computed in Ref.~\cite{Berwein:2015vca}. We show the results thus obtained in Fig.~\ref{fg:bbg_Cheung}. For the bottom we have used the RS$^\prime$ mass $m_b^{\rm RS}(\textrm{1 GeV})=4.863$~GeV.


\noindent\emph{Conclusions}.\,---\,Using a recently developed nonrelativistic effective field theory~\cite{Berwein:2015vca,Oncala:2017hop,Brambilla:2017uyf} we have obtained for the first time the $1/m$ and $1/m^2$ spin interactions characterizing the fine and hyperfine splittings of heavy quarkonium hybrids, one type of exotic hadron under intensive study at high-energy experiments at the B-factory in Japan, the $\tau$-charm factory in China as well at the LHC at CERN and in perspective at FAIR in Germany.

The spin interactions in quarkonium hybrids display novel features. The most interesting one is the appearance of operators already at order $1/m$ that couple the total spin of the heavy-quark-antiquark pair with the spin of the gluons~\cite{Oncala:2017hop}. At order $1/m^2$, we have the spin-orbit, spin-spin and tensor operators familiar from the studies of standard quarkonia. In addition, at order $1/m^2$, three new relevant operators appear involving the projectors associated to the representations of $D_{\infty h}$.

The spin interactions depend on some matching coefficients. Their contributions coming from integrating out the heavy-quark mass and the heavy-quark-antiquark distance are of perturbative nature, and, therefore, can be calculated in an expansion in the strong coupling. The contributions coming from integrating out the scale $\Lambda_{\rm QCD}$ are nonperturbative. Due to the separation of scales underlying the construction of the EFT, the heavy quark-antiquark flavor dependence of the matching coefficients can be factorized from the gluon correlators encoding the nonperturbative dynamics. Thus nonperturbative contributions to the matching coefficients can be determined in the charm sector and then used in the bottom sector. Using the latest lattice-QCD calculations of the charmonium hybrid spectrum we have determined the nonperturbative contributions and obtained new predictions for the bottomonium hybrid spin splittings where lattice-QCD calculations are still very challenging and incomplete.

We have found that the perturbative contributions, corresponding to the spin-dependent color-octet quark-antiquark potentials, generate a spin-splitting pattern opposite to the one observed in lattice QCD. In our framework, this deviation is compensated by the nonperturbative contributions, in particular, by the order $1/m$ operator that is peculiar of hybrid states and has no perturbative counterpart.

Finally, we note that, since this EFT setup can be generalized to describe states with light degrees of freedom other than gluons, such as heavy tetraquarks and pentaquarks~\cite{Braaten:2014ita,Brambilla:2017uyf}, a similar analysis could be done to describe spin multiplets of exotic quarkonia other than hybrids, eventually providing an unified description of all heavy-quark-antiquark spin multiplets.

\begin{acknowledgments}
\noindent\emph{Acknowledgments}.\,---\,This work has been supported by the DFG and the NSFC through funds provided to the Sino-German CRC 110 ``Symmetries and the emergence of structure in QCD'', and by the DFG cluster of excellence ``Origin and structure of the Universe'' (www.universe-cluster.de). J.T.C. has been supported by the Spanish MINECO's Grants No.\ FPA2014-55613-P, No.\ FPA2017-86989-P and SEV-2016-0588. J.S. acknowledges the financial support from the Alexander von Humboldt Foundation, from the European Union's Horizon 2020 research and innovation program under the Marie Sk\l{}odowska--Curie Grant No.\ 665919, and from Spanish MINECO's Juan de la Cierva-Incorporaci\'on program, Grant Agreement No. IJCI-2016-30028.
\end{acknowledgments}

\appendix

\section{Matrix elements of operators involving \texorpdfstring{$\bm{L}_{Q\bar{Q}}$}{LQQb}} \label{appendix}

The angular momentum operator in spherical coordinates reads
\begin{align}
\bm{L}_{Q\bar{Q}}=-i\hat{\bm{\phi}}\pa_{\theta}+\frac{i}{\sin\theta}\hat{\bm{\theta}}\pa_{\phi}\,.
\end{align}
One can compute the following commutators of the angular momentum operator and the unit vectors in spherical coordinates
\begin{align}
\left[{L}_{Q\bar{Q}}^i,\hat{r}_{0}^j\right]&=\hat{r}^i_+\hat{r}^j_--\hat{r}^i_-\hat{r}^j_+\,, \\
\left[{L}_{Q\bar{Q}}^i,\hat{\theta}^j\right]&=i \hat{\phi}^i\hat{r}_0^j+i\cot(\theta)\hat{\theta}^i\hat{\phi}^j\,, \\
\left[{L}_{Q\bar{Q}}^i,\hat{\phi}^j\right]&=-i\hat{\theta}^i\left(\hat{r}_0^j+\cot(\theta)\hat{\theta}^j\right)\,,
\end{align}
from which we obtain the commutators with the projection vectors
\begin{align}
\left[{L}_{Q\bar{Q}}^i,\hat{r}_{\pm}^j\right]=\pm\left(\hat{r}^j_0\hat{r}^i_{\pm}+\cot(\theta)\hat{\theta}^i\hat{r}^j_{\pm}\right)\,.
\end{align}
For any $\la$, we have
\begin{align}
\left[{L}_{Q\bar{Q}}^i,\hat{r}_{\la}^j\right]&=\la\cot(\theta)\hat{r}_{\la}^j\hat{\theta}^i+\sqrt{1-\frac{\la(\la-1)}{2}}\hat{r}^j_{\la-1}\hat{r}^i_{+}\nonumber\\
&-\sqrt{1-\frac{\la(\la+1)}{2}}\hat{r}^j_{\la+1}\hat{r}^i_{-}\,.
\end{align}
To compute the matrix elements of $V_{SLa}$ we rewrite the operator in the following way,
\begin{align}
&\left(\hat{r}_{\la}^{i\dag}\bm{L}_{Q\bar{Q}}\hat{r}^{i}_{\lap}\right)\cdot \bm{S}=\left(\bm{L}_{Q\bar{Q}}\de_{\la\lap}+\hat{r}_{\la}^{i\dag}\left[\bm{L}_{Q\bar{Q}},\,\hat{r}^{i}_{\lap}\right]\right)\cdot \bm{S}\nonumber\\
&=\left(
\begin{array}{ccc}
 \bm{L}_{Q\bar{Q}} & \hat{\bm{r}}_{+}^{\dag} & -\hat{\bm{r}}_{-}^{\dag} \\
 \hat{\bm{r}}_{+}  & \bm{L}_{Q\bar{Q}}+\cot\theta \hat{\bm{\theta}} &  0 \\
 -\hat{\bm{r}}_{-} & 0 & \bm{L}_{Q\bar{Q}}-\cot\theta \hat{\bm{\theta}} \\
\end{array}\right)\cdot \bm{S}\nonumber\\
&=\de_{\la\lap}\left[\bm{L}_{Q\bar{Q}}+\la\left(\cot\theta \hat{\bm{\theta}}+\hat{\bm{r}}_0\right)\right]\cdot\bm{S}+i\left(\hat{\bm{r}}^{\dag}_{\la}\times \hat{\bm{r}}_{\lap}\right)\cdot\bm{S}\nonumber\\
&=\de_{\la\lap}\bm{L}\cdot\bm{S}-\left(\hat{r}_{\la}^{i\dag}(\bm{K}_1)^{ij}\hat{r}^{j}_{\lap}\right)\cdot \bm{S}\nonumber\\
&=\frac{\de_{\la\lap}}{2}\left(\bm{J}^2-\bm{L}^2-\bm{S}^2\right)-\left(\hat{r}_{\la}^{i\dag}(\bm{K}_1)^{ij}\hat{r}^{j}_{\lap}\right)\cdot \bm{S}\,,
\end{align}
where we have used that
\begin{align}
&\left[\bm{L}_{Q\bar{Q}}+\la\left(\cot\theta \hat{\bm{\theta}}+\hat{\bm{r}}_0\right)\right]^2\nonumber\\
&=\bm{L}^2_{Q\bar{Q}}+\frac{\la^2}{\sin^2\theta}+2i\la\frac{\cos\theta}{\sin^2\theta}\partial_{\phi}\equiv\bm{L}^2\,,
\end{align}
which is the operator whose eigenfunctions are our angular wave functions,
\begin{align}
\left(\bm{L}^2_{Q\bar{Q}}+\frac{\la^2}{\sin^2\theta}+2i\la\frac{\cos\theta}{\sin^2\theta}\partial_{\phi}\right)v^{\la}_{lm}(\theta,\phi)=l(l+1)v^{\la}_{lm}(\theta,\phi)\,.
\end{align}
Next we show a detailed computation of the matrix elements of the operator $V_{SLb}$,
\begin{align}
\hat{r}_{\lap}^{\dag i}\left({L}_{Q\bar{Q}}^i{S}^l+{S}^i{L}_{Q\bar{Q}}^l\right)\hat{r}_{\la}^l\,.\label{eq:VSLb_op}
\end{align}
The first term in Eq.~\eqref{eq:VSLb_op} can be manipulated as follows:
\begin{align}
\hat{r}_{\la'}^{\dag i}{S}^i{L}_{Q\bar{Q}}^l\hat{r}_{\la}^l&=(\hat{\bm r}_{\la'}^{\dag}\cdot\bm{S})(\hat{\bm r}_{\la}\cdot\bm{L}_{Q\bar{Q}}+[{L}_{Q\bar{Q}}^l,\,\hat{ r}_{\la}^l])\nonumber\\
&=(\hat{\bm r}_{\la'}^{\dag}\cdot\bm{S})\left(\hat{\bm r}_{\la}\cdot\bm{L}_{Q\bar{Q}}-\lambda^2\frac{\cot(\theta)}{\sqrt{2}}\right)\,.
\end{align}
This expression vanishes for $\lambda=0$. In the case $\lambda=\pm1$,
\begin{align}
\hat{r}_{\la'}^{\dag i}{S}^i{L}_{Q\bar{Q}}^l\hat{r}_{\pm}^l&=(\hat{\bm r}_{\la'}^{\dag}\cdot\bm{S})\left(\hat{\bm r}_{\pm}\cdot\bm{L}_{Q\bar{Q}}-\frac{\cot(\theta)}{\sqrt{2}}\right)\nonumber\\
&=\mp\frac{(\hat{\bm r}_{\la'}^{\dag}\cdot\bm{S})}{\sqrt{2}}\left(\pm\pa_{\theta}+\frac{i}{\sin(\theta)}\pa_{\phi}+\la\cot(\theta)\right)\nonumber \\
&=\mp\frac{(\hat{\bm r}_{\la'}^{\dag}\cdot\bm{S})}{\sqrt{2}}\mathcal{K}_{\mp}\,.
\end{align}
The operators $\mathcal{K}_{\pm}$ act as the $\la$-raising and -lowering operators for the angular wave functions $v^{\la}_{l m_l}$,
\begin{align}
\mathcal{K}_{\pm}v^{\la}_{l m_l}(\theta,\phi)=\sqrt{l(l+1)-\la(\la\pm 1)}v^{\la \pm 1}_{l m_l}(\theta,\phi)\,.
\end{align}
The second piece of the operator in Eq.~(\ref{eq:VSLb_op}) can be written in a similar way,
\begin{align}
\hat{r}_{\la'}^{\dag i}{L}_{Q\bar{Q}}^i{S}^l\hat{r}_{\la}^l&=\left([\hat{r}_{\la'}^{\dag i},{L}^i]+\bm{L}_{Q\bar{Q}}\cdot\hat{\bm r}_{\la'}^{\dag}\right)\bm{S}\cdot\hat{\bm r}_{\la}\nonumber\\
&=\left(\bm{L}_{Q\bar{Q}}\cdot\hat{\bm r}_{\la'}^{\dag}-(\lap)^2\frac{\cot(\theta)}{\sqrt{2}}\right)\hat{\bm r}_{\la}\cdot\bm{S}\,.
\end{align}
In this case the operator vanishes for $\lap=0$. For $\lap=\pm1$ we have
\begin{align}
\hat{r}_{\pm}^{\dag i}{L}_{Q\bar{Q}}^i{S}^l\hat{r}_{\la}^l&=\left(\left(\hat{\bm r}_{\pm}\cdot\bm{L}_{Q\bar{Q}}\right)^{\dag}-\frac{\cot(\theta)}{\sqrt{2}}\right)\hat{\bm r}_{\la}\cdot\bm{S}\nonumber\\
&=\mp\left(\pm\pa_{\theta}+\frac{i}{\sin(\theta)}\pa_{\phi}+\lap\cot(\theta)\right)^{\dag}\frac{\hat{\bm r}_{\la}\cdot\bm{S}}{\sqrt{2}}\nonumber \\
&=\mp \mathcal{K}^{\prime\dag}_{\mp}\frac{(\hat{\bm r}_{\la}\cdot\bm{S})}{\sqrt{2}}\,.
\end{align}
The prime in $\mathcal{K}^{\prime}_{\mp}$ indicates that the operator depends on $\lap$ instead of $\la$. Adding up both contributions we arrive at
\begin{align}
&\hat{r}_{\la}^{\dag i}\left({L}_{Q\bar{Q}}^i{S}^l+{S}^i{L}_{Q\bar{Q}}^l\right)\hat{r}_{\lap}^l \nonumber\\
&=\mp \mathcal{K}^{\dag}_{\mp}\frac{(\hat{\bm r}_{\lap}\cdot\bm{S})}{\sqrt{2}}\de_{\la\pm 1}
\mp\frac{(\hat{\bm r}_{\la}^{\dag}\cdot\bm{S})}{\sqrt{2}}\mathcal{K}^{\prime}_{\mp}\de_{\lap\pm 1}\,.
\end{align}


\bibliography{short_SpinHybrids}

\begin{thebibliography}{55}%
\makeatletter
\providecommand \@ifxundefined [1]{%
 \@ifx{#1\undefined}
}%
\providecommand \@ifnum [1]{%
 \ifnum #1\expandafter \@firstoftwo
 \else \expandafter \@secondoftwo
 \fi
}%
\providecommand \@ifx [1]{%
 \ifx #1\expandafter \@firstoftwo
 \else \expandafter \@secondoftwo
 \fi
}%
\providecommand \natexlab [1]{#1}%
\providecommand \enquote  [1]{``#1''}%
\providecommand \bibnamefont  [1]{#1}%
\providecommand \bibfnamefont [1]{#1}%
\providecommand \citenamefont [1]{#1}%
\providecommand \href@noop [0]{\@secondoftwo}%
\providecommand \href [0]{\begingroup \@sanitize@url \@href}%
\providecommand \@href[1]{\@@startlink{#1}\@@href}%
\providecommand \@@href[1]{\endgroup#1\@@endlink}%
\providecommand \@sanitize@url [0]{\catcode `\\12\catcode `\$12\catcode
  `\&12\catcode `\#12\catcode `\^12\catcode `\_12\catcode `\%12\relax}%
\providecommand \@@startlink[1]{}%
\providecommand \@@endlink[0]{}%
\providecommand \url  [0]{\begingroup\@sanitize@url \@url }%
\providecommand \@url [1]{\endgroup\@href {#1}{\urlprefix }}%
\providecommand \urlprefix  [0]{URL }%
\providecommand \Eprint [0]{\href }%
\providecommand \doibase [0]{http://dx.doi.org/}%
\providecommand \selectlanguage [0]{\@gobble}%
\providecommand \bibinfo  [0]{\@secondoftwo}%
\providecommand \bibfield  [0]{\@secondoftwo}%
\providecommand \translation [1]{[#1]}%
\providecommand \BibitemOpen [0]{}%
\providecommand \bibitemStop [0]{}%
\providecommand \bibitemNoStop [0]{.\EOS\space}%
\providecommand \EOS [0]{\spacefactor3000\relax}%
\providecommand \BibitemShut  [1]{\csname bibitem#1\endcsname}%
\let\auto@bib@innerbib\@empty
\bibitem [{\citenamefont {Gell-Mann}(1964)}]{GellMann:1964nj}%
  \BibitemOpen
  \bibfield  {author} {\bibinfo {author} {\bibfnamefont {M.}~\bibnamefont
  {Gell-Mann}},\ }\href {\doibase 10.1016/S0031-9163(64)92001-3} {\bibfield
  {journal} {\bibinfo  {journal} {Phys. Lett.}\ }\textbf {\bibinfo {volume}
  {8}},\ \bibinfo {pages} {214} (\bibinfo {year} {1964})}\BibitemShut {NoStop}%
\bibitem [{\citenamefont {Zweig}(1964)}]{Zweig:1964CERN}%
  \BibitemOpen
  \bibfield  {author} {\bibinfo {author} {\bibfnamefont {G.}~\bibnamefont
  {Zweig}},\ }\href@noop {} {\bibfield  {journal} {\bibinfo  {journal} {CERN
  Report No.8182/TH.401, CERN Report No.8419/TH.412}\ } (\bibinfo {year}
  {1964})}\BibitemShut {NoStop}%
\bibitem [{\citenamefont {Choi}\ \emph {et~al.}(2003)\citenamefont {Choi} \emph
  {et~al.}}]{Choi:2003ue}%
  \BibitemOpen
  \bibfield  {author} {\bibinfo {author} {\bibfnamefont {S.~K.}\ \bibnamefont
  {Choi}} \emph {et~al.} (\bibinfo {collaboration} {Belle}),\ }\href {\doibase
  10.1103/PhysRevLett.91.262001} {\bibfield  {journal} {\bibinfo  {journal}
  {Phys. Rev. Lett.}\ }\textbf {\bibinfo {volume} {91}},\ \bibinfo {pages}
  {262001} (\bibinfo {year} {2003})},\ \Eprint
  {http://arxiv.org/abs/hep-ex/0309032} {arXiv:hep-ex/0309032 [hep-ex]}
  \BibitemShut {NoStop}%
\bibitem [{\citenamefont {Brambilla}\ \emph
  {et~al.}(2005{\natexlab{a}})\citenamefont {Brambilla} \emph
  {et~al.}}]{Brambilla:2004wf}%
  \BibitemOpen
  \bibfield  {author} {\bibinfo {author} {\bibfnamefont {N.}~\bibnamefont
  {Brambilla}} \emph {et~al.},\ }\href@noop {} {\bibfield  {journal} {\bibinfo
  {journal} {{CERN-2005-005 (CERN, Geneva,2005), arXiv:hep-ph/0412158
  [hep-ph]}}\ } (\bibinfo {year} {2005}{\natexlab{a}})},\ \Eprint
  {http://arxiv.org/abs/hep-ph/0412158} {arXiv:hep-ph/0412158 [hep-ph]}
  \BibitemShut {NoStop}%
\bibitem [{\citenamefont {Brambilla}\ \emph {et~al.}(2011)\citenamefont
  {Brambilla} \emph {et~al.}}]{Brambilla:2010cs}%
  \BibitemOpen
  \bibfield  {author} {\bibinfo {author} {\bibfnamefont {N.}~\bibnamefont
  {Brambilla}} \emph {et~al.},\ }\href {\doibase
  10.1140/epjc/s10052-010-1534-9} {\bibfield  {journal} {\bibinfo  {journal}
  {Eur. Phys. J.}\ }\textbf {\bibinfo {volume} {C71}},\ \bibinfo {pages} {1534}
  (\bibinfo {year} {2011})},\ \Eprint {http://arxiv.org/abs/1010.5827}
  {arXiv:1010.5827 [hep-ph]} \BibitemShut {NoStop}%
\bibitem [{\citenamefont {Brambilla}\ \emph {et~al.}(2014)\citenamefont
  {Brambilla} \emph {et~al.}}]{Brambilla:2014jmp}%
  \BibitemOpen
  \bibfield  {author} {\bibinfo {author} {\bibfnamefont {N.}~\bibnamefont
  {Brambilla}} \emph {et~al.},\ }\href {\doibase
  10.1140/epjc/s10052-014-2981-5} {\bibfield  {journal} {\bibinfo  {journal}
  {Eur. Phys. J.}\ }\textbf {\bibinfo {volume} {C74}},\ \bibinfo {pages} {2981}
  (\bibinfo {year} {2014})},\ \Eprint {http://arxiv.org/abs/1404.3723}
  {arXiv:1404.3723 [hep-ph]} \BibitemShut {NoStop}%
\bibitem [{\citenamefont {Olsen}(2015)}]{Olsen:2014qna}%
  \BibitemOpen
  \bibfield  {author} {\bibinfo {author} {\bibfnamefont {S.~L.}\ \bibnamefont
  {Olsen}},\ }\href {\doibase 10.1007/S11467-014-0449-6} {\bibfield  {journal}
  {\bibinfo  {journal} {Front. Phys.(Beijing)}\ }\textbf {\bibinfo {volume}
  {10}},\ \bibinfo {pages} {121} (\bibinfo {year} {2015})},\ \Eprint
  {http://arxiv.org/abs/1411.7738} {arXiv:1411.7738 [hep-ex]} \BibitemShut
  {NoStop}%
\bibitem [{\citenamefont {Lutz}\ \emph {et~al.}(2009)\citenamefont {Lutz} \emph
  {et~al.}}]{Lutz:2009ff}%
  \BibitemOpen
  \bibfield  {author} {\bibinfo {author} {\bibfnamefont {M.~F.~M.}\
  \bibnamefont {Lutz}} \emph {et~al.} (\bibinfo {collaboration} {PANDA}),\
  }\href@noop {} {\  (\bibinfo {year} {2009})},\ \bibinfo {note}
  {arXiv:hep-ex/0903.3905}\BibitemShut {NoStop}%
\bibitem [{\citenamefont {Dudek}\ \emph {et~al.}(2012)\citenamefont {Dudek}
  \emph {et~al.}}]{Dudek:2012vr}%
  \BibitemOpen
  \bibfield  {author} {\bibinfo {author} {\bibfnamefont {J.}~\bibnamefont
  {Dudek}} \emph {et~al.},\ }\href {\doibase 10.1140/epja/i2012-12187-1}
  {\bibfield  {journal} {\bibinfo  {journal} {Eur. Phys. J.}\ }\textbf
  {\bibinfo {volume} {A48}},\ \bibinfo {pages} {187} (\bibinfo {year}
  {2012})},\ \Eprint {http://arxiv.org/abs/1208.1244} {arXiv:1208.1244
  [hep-ex]} \BibitemShut {NoStop}%
\bibitem [{\citenamefont {Horn}\ and\ \citenamefont
  {Mandula}(1978)}]{Horn:1977rq}%
  \BibitemOpen
  \bibfield  {author} {\bibinfo {author} {\bibfnamefont {D.}~\bibnamefont
  {Horn}}\ and\ \bibinfo {author} {\bibfnamefont {J.}~\bibnamefont {Mandula}},\
  }\href {\doibase 10.1103/PhysRevD.17.898} {\bibfield  {journal} {\bibinfo
  {journal} {Phys. Rev.}\ }\textbf {\bibinfo {volume} {D17}},\ \bibinfo {pages}
  {898} (\bibinfo {year} {1978})}\BibitemShut {NoStop}%
\bibitem [{\citenamefont {Barnes}(1981)}]{Barnes:1981ac}%
  \BibitemOpen
  \bibfield  {author} {\bibinfo {author} {\bibfnamefont {T.}~\bibnamefont
  {Barnes}},\ }\href {\doibase 10.1007/BF01549736} {\bibfield  {journal}
  {\bibinfo  {journal} {Z. Phys.}\ }\textbf {\bibinfo {volume} {C10}},\
  \bibinfo {pages} {275} (\bibinfo {year} {1981})}\BibitemShut {NoStop}%
\bibitem [{\citenamefont {Chanowitz}\ and\ \citenamefont
  {Sharpe}(1983)}]{Chanowitz:1982qj}%
  \BibitemOpen
  \bibfield  {author} {\bibinfo {author} {\bibfnamefont {M.~S.}\ \bibnamefont
  {Chanowitz}}\ and\ \bibinfo {author} {\bibfnamefont {S.~R.}\ \bibnamefont
  {Sharpe}},\ }\href {\doibase 10.1016/0550-3213(83)90561-8,
  10.1016/0550-3213(83)90635-1} {\bibfield  {journal} {\bibinfo  {journal}
  {Nucl. Phys.}\ }\textbf {\bibinfo {volume} {B222}},\ \bibinfo {pages} {211}
  (\bibinfo {year} {1983})},\ \bibinfo {note} {[Erratum: Nucl. Phys. B228, 588
  (1983)]}\BibitemShut {NoStop}%
\bibitem [{\citenamefont {Barnes}\ \emph {et~al.}(1983)\citenamefont {Barnes},
  \citenamefont {Close}, \citenamefont {de~Viron},\ and\ \citenamefont
  {Weyers}}]{Barnes:1982tx}%
  \BibitemOpen
  \bibfield  {author} {\bibinfo {author} {\bibfnamefont {T.}~\bibnamefont
  {Barnes}}, \bibinfo {author} {\bibfnamefont {F.~E.}\ \bibnamefont {Close}},
  \bibinfo {author} {\bibfnamefont {F.}~\bibnamefont {de~Viron}}, \ and\
  \bibinfo {author} {\bibfnamefont {J.}~\bibnamefont {Weyers}},\ }\href
  {\doibase 10.1016/0550-3213(83)90004-4} {\bibfield  {journal} {\bibinfo
  {journal} {Nucl. Phys.}\ }\textbf {\bibinfo {volume} {B224}},\ \bibinfo
  {pages} {241} (\bibinfo {year} {1983})}\BibitemShut {NoStop}%
\bibitem [{\citenamefont {Cornwall}\ and\ \citenamefont
  {Soni}(1983)}]{Cornwall:1982zn}%
  \BibitemOpen
  \bibfield  {author} {\bibinfo {author} {\bibfnamefont {J.~M.}\ \bibnamefont
  {Cornwall}}\ and\ \bibinfo {author} {\bibfnamefont {A.}~\bibnamefont
  {Soni}},\ }\href {\doibase 10.1016/0370-2693(83)90481-1} {\bibfield
  {journal} {\bibinfo  {journal} {Phys. Lett.}\ }\textbf {\bibinfo {volume}
  {B120}},\ \bibinfo {pages} {431} (\bibinfo {year} {1983})}\BibitemShut
  {NoStop}%
\bibitem [{\citenamefont {Isgur}\ and\ \citenamefont
  {Paton}(1985)}]{Isgur:1984bm}%
  \BibitemOpen
  \bibfield  {author} {\bibinfo {author} {\bibfnamefont {N.}~\bibnamefont
  {Isgur}}\ and\ \bibinfo {author} {\bibfnamefont {J.~E.}\ \bibnamefont
  {Paton}},\ }\href {\doibase 10.1103/PhysRevD.31.2910} {\bibfield  {journal}
  {\bibinfo  {journal} {Phys. Rev.}\ }\textbf {\bibinfo {volume} {D31}},\
  \bibinfo {pages} {2910} (\bibinfo {year} {1985})}\BibitemShut {NoStop}%
\bibitem [{\citenamefont {Kalashnikova}(1994)}]{Kalashnikova:1993xb}%
  \BibitemOpen
  \bibfield  {author} {\bibinfo {author} {\bibfnamefont {{\relax Yu}.~S.}\
  \bibnamefont {Kalashnikova}},\ }\href {\doibase 10.1007/BF01560246}
  {\bibfield  {journal} {\bibinfo  {journal} {Z. Phys.}\ }\textbf {\bibinfo
  {volume} {C62}},\ \bibinfo {pages} {323} (\bibinfo {year}
  {1994})}\BibitemShut {NoStop}%
\bibitem [{\citenamefont {Swanson}\ and\ \citenamefont
  {Szczepaniak}(1998)}]{Swanson:1998kx}%
  \BibitemOpen
  \bibfield  {author} {\bibinfo {author} {\bibfnamefont {E.~S.}\ \bibnamefont
  {Swanson}}\ and\ \bibinfo {author} {\bibfnamefont {A.~P.}\ \bibnamefont
  {Szczepaniak}},\ }\href {\doibase 10.1103/PhysRevD.59.014035} {\bibfield
  {journal} {\bibinfo  {journal} {Phys. Rev.}\ }\textbf {\bibinfo {volume}
  {D59}},\ \bibinfo {pages} {014035} (\bibinfo {year} {1998})},\ \Eprint
  {http://arxiv.org/abs/hep-ph/9804219} {arXiv:hep-ph/9804219 [hep-ph]}
  \BibitemShut {NoStop}%
\bibitem [{\citenamefont {Brau}\ and\ \citenamefont
  {Semay}(2004)}]{Brau:2004xw}%
  \BibitemOpen
  \bibfield  {author} {\bibinfo {author} {\bibfnamefont {F.}~\bibnamefont
  {Brau}}\ and\ \bibinfo {author} {\bibfnamefont {C.}~\bibnamefont {Semay}},\
  }\href {\doibase 10.1103/PhysRevD.70.014017} {\bibfield  {journal} {\bibinfo
  {journal} {Phys. Rev.}\ }\textbf {\bibinfo {volume} {D70}},\ \bibinfo {pages}
  {014017} (\bibinfo {year} {2004})},\ \Eprint
  {http://arxiv.org/abs/hep-ph/0412173} {arXiv:hep-ph/0412173 [hep-ph]}
  \BibitemShut {NoStop}%
\bibitem [{\citenamefont {Szczepaniak}\ and\ \citenamefont
  {Krupinski}(2006)}]{Szczepaniak:2005xi}%
  \BibitemOpen
  \bibfield  {author} {\bibinfo {author} {\bibfnamefont {A.~P.}\ \bibnamefont
  {Szczepaniak}}\ and\ \bibinfo {author} {\bibfnamefont {P.}~\bibnamefont
  {Krupinski}},\ }\href {\doibase 10.1103/PhysRevD.73.034022} {\bibfield
  {journal} {\bibinfo  {journal} {Phys. Rev.}\ }\textbf {\bibinfo {volume}
  {D73}},\ \bibinfo {pages} {034022} (\bibinfo {year} {2006})},\ \Eprint
  {http://arxiv.org/abs/hep-ph/0511083} {arXiv:hep-ph/0511083 [hep-ph]}
  \BibitemShut {NoStop}%
\bibitem [{\citenamefont {Buisseret}\ and\ \citenamefont
  {Semay}(2006)}]{Buisseret:2006wc}%
  \BibitemOpen
  \bibfield  {author} {\bibinfo {author} {\bibfnamefont {F.}~\bibnamefont
  {Buisseret}}\ and\ \bibinfo {author} {\bibfnamefont {C.}~\bibnamefont
  {Semay}},\ }\href {\doibase 10.1103/PhysRevD.74.114018} {\bibfield  {journal}
  {\bibinfo  {journal} {Phys. Rev.}\ }\textbf {\bibinfo {volume} {D74}},\
  \bibinfo {pages} {114018} (\bibinfo {year} {2006})},\ \Eprint
  {http://arxiv.org/abs/hep-ph/0610132} {arXiv:hep-ph/0610132 [hep-ph]}
  \BibitemShut {NoStop}%
\bibitem [{\citenamefont {Guo}\ \emph {et~al.}(2008)\citenamefont {Guo},
  \citenamefont {Szczepaniak}, \citenamefont {Galata}, \citenamefont
  {Vassallo},\ and\ \citenamefont {Santopinto}}]{Guo:2008yz}%
  \BibitemOpen
  \bibfield  {author} {\bibinfo {author} {\bibfnamefont {P.}~\bibnamefont
  {Guo}}, \bibinfo {author} {\bibfnamefont {A.~P.}\ \bibnamefont
  {Szczepaniak}}, \bibinfo {author} {\bibfnamefont {G.}~\bibnamefont {Galata}},
  \bibinfo {author} {\bibfnamefont {A.}~\bibnamefont {Vassallo}}, \ and\
  \bibinfo {author} {\bibfnamefont {E.}~\bibnamefont {Santopinto}},\ }\href
  {\doibase 10.1103/PhysRevD.78.056003} {\bibfield  {journal} {\bibinfo
  {journal} {Phys. Rev.}\ }\textbf {\bibinfo {volume} {D78}},\ \bibinfo {pages}
  {056003} (\bibinfo {year} {2008})},\ \Eprint {http://arxiv.org/abs/0807.2721}
  {arXiv:0807.2721 [hep-ph]} \BibitemShut {NoStop}%
\bibitem [{\citenamefont {Govaerts}\ \emph
  {et~al.}(1985{\natexlab{a}})\citenamefont {Govaerts}, \citenamefont
  {Reinders}, \citenamefont {Rubinstein},\ and\ \citenamefont
  {Weyers}}]{Govaerts:1984hc}%
  \BibitemOpen
  \bibfield  {author} {\bibinfo {author} {\bibfnamefont {J.}~\bibnamefont
  {Govaerts}}, \bibinfo {author} {\bibfnamefont {L.~J.}\ \bibnamefont
  {Reinders}}, \bibinfo {author} {\bibfnamefont {H.~R.}\ \bibnamefont
  {Rubinstein}}, \ and\ \bibinfo {author} {\bibfnamefont {J.}~\bibnamefont
  {Weyers}},\ }\href {\doibase 10.1016/0550-3213(85)90609-1} {\bibfield
  {journal} {\bibinfo  {journal} {Nucl. Phys.}\ }\textbf {\bibinfo {volume}
  {B258}},\ \bibinfo {pages} {215} (\bibinfo {year}
  {1985}{\natexlab{a}})}\BibitemShut {NoStop}%
\bibitem [{\citenamefont {Govaerts}\ \emph
  {et~al.}(1985{\natexlab{b}})\citenamefont {Govaerts}, \citenamefont
  {Reinders},\ and\ \citenamefont {Weyers}}]{Govaerts:1985fx}%
  \BibitemOpen
  \bibfield  {author} {\bibinfo {author} {\bibfnamefont {J.}~\bibnamefont
  {Govaerts}}, \bibinfo {author} {\bibfnamefont {L.~J.}\ \bibnamefont
  {Reinders}}, \ and\ \bibinfo {author} {\bibfnamefont {J.}~\bibnamefont
  {Weyers}},\ }\href {\doibase 10.1016/0550-3213(85)90505-X} {\bibfield
  {journal} {\bibinfo  {journal} {Nucl. Phys.}\ }\textbf {\bibinfo {volume}
  {B262}},\ \bibinfo {pages} {575} (\bibinfo {year}
  {1985}{\natexlab{b}})}\BibitemShut {NoStop}%
\bibitem [{\citenamefont {Govaerts}\ \emph {et~al.}(1987)\citenamefont
  {Govaerts}, \citenamefont {Reinders}, \citenamefont {Francken}, \citenamefont
  {Gonze},\ and\ \citenamefont {Weyers}}]{Govaerts:1986pp}%
  \BibitemOpen
  \bibfield  {author} {\bibinfo {author} {\bibfnamefont {J.}~\bibnamefont
  {Govaerts}}, \bibinfo {author} {\bibfnamefont {L.~J.}\ \bibnamefont
  {Reinders}}, \bibinfo {author} {\bibfnamefont {P.}~\bibnamefont {Francken}},
  \bibinfo {author} {\bibfnamefont {X.}~\bibnamefont {Gonze}}, \ and\ \bibinfo
  {author} {\bibfnamefont {J.}~\bibnamefont {Weyers}},\ }\href {\doibase
  10.1016/0550-3213(87)90056-3} {\bibfield  {journal} {\bibinfo  {journal}
  {Nucl. Phys.}\ }\textbf {\bibinfo {volume} {B284}},\ \bibinfo {pages} {674}
  (\bibinfo {year} {1987})}\BibitemShut {NoStop}%
\bibitem [{\citenamefont {Harnett}\ \emph {et~al.}(2012)\citenamefont
  {Harnett}, \citenamefont {Kleiv}, \citenamefont {Steele},\ and\ \citenamefont
  {Jin}}]{Harnett:2012gs}%
  \BibitemOpen
  \bibfield  {author} {\bibinfo {author} {\bibfnamefont {D.}~\bibnamefont
  {Harnett}}, \bibinfo {author} {\bibfnamefont {R.~T.}\ \bibnamefont {Kleiv}},
  \bibinfo {author} {\bibfnamefont {T.~G.}\ \bibnamefont {Steele}}, \ and\
  \bibinfo {author} {\bibfnamefont {H.-y.}\ \bibnamefont {Jin}},\ }\href
  {\doibase 10.1088/0954-3899/39/12/125003} {\bibfield  {journal} {\bibinfo
  {journal} {J. Phys.}\ }\textbf {\bibinfo {volume} {G39}},\ \bibinfo {pages}
  {125003} (\bibinfo {year} {2012})},\ \Eprint {http://arxiv.org/abs/1206.6776}
  {arXiv:1206.6776 [hep-ph]} \BibitemShut {NoStop}%
\bibitem [{\citenamefont {Berg}\ \emph {et~al.}(2012)\citenamefont {Berg},
  \citenamefont {Harnett}, \citenamefont {Kleiv},\ and\ \citenamefont
  {Steele}}]{Berg:2012gd}%
  \BibitemOpen
  \bibfield  {author} {\bibinfo {author} {\bibfnamefont {R.}~\bibnamefont
  {Berg}}, \bibinfo {author} {\bibfnamefont {D.}~\bibnamefont {Harnett}},
  \bibinfo {author} {\bibfnamefont {R.~T.}\ \bibnamefont {Kleiv}}, \ and\
  \bibinfo {author} {\bibfnamefont {T.~G.}\ \bibnamefont {Steele}},\ }\href
  {\doibase 10.1103/PhysRevD.86.034002} {\bibfield  {journal} {\bibinfo
  {journal} {Phys. Rev.}\ }\textbf {\bibinfo {volume} {D86}},\ \bibinfo {pages}
  {034002} (\bibinfo {year} {2012})},\ \Eprint {http://arxiv.org/abs/1204.0049}
  {arXiv:1204.0049 [hep-ph]} \BibitemShut {NoStop}%
\bibitem [{\citenamefont {Chen}\ \emph {et~al.}(2013)\citenamefont {Chen},
  \citenamefont {Kleiv}, \citenamefont {Steele}, \citenamefont {Bulthuis},
  \citenamefont {Harnett}, \citenamefont {Ho}, \citenamefont {Richards},\ and\
  \citenamefont {Zhu}}]{Chen:2013zia}%
  \BibitemOpen
  \bibfield  {author} {\bibinfo {author} {\bibfnamefont {W.}~\bibnamefont
  {Chen}}, \bibinfo {author} {\bibfnamefont {R.~T.}\ \bibnamefont {Kleiv}},
  \bibinfo {author} {\bibfnamefont {T.~G.}\ \bibnamefont {Steele}}, \bibinfo
  {author} {\bibfnamefont {B.}~\bibnamefont {Bulthuis}}, \bibinfo {author}
  {\bibfnamefont {D.}~\bibnamefont {Harnett}}, \bibinfo {author} {\bibfnamefont
  {J.}~\bibnamefont {Ho}}, \bibinfo {author} {\bibfnamefont {T.}~\bibnamefont
  {Richards}}, \ and\ \bibinfo {author} {\bibfnamefont {S.-L.}\ \bibnamefont
  {Zhu}},\ }\href {\doibase 10.1007/JHEP09(2013)019} {\bibfield  {journal}
  {\bibinfo  {journal} {JHEP}\ }\textbf {\bibinfo {volume} {09}},\ \bibinfo
  {pages} {019} (\bibinfo {year} {2013})},\ \Eprint
  {http://arxiv.org/abs/1304.4522} {arXiv:1304.4522 [hep-ph]} \BibitemShut
  {NoStop}%
\bibitem [{\citenamefont {Lebed}\ \emph {et~al.}(2017)\citenamefont {Lebed},
  \citenamefont {Mitchell},\ and\ \citenamefont {Swanson}}]{Lebed:2016hpi}%
  \BibitemOpen
  \bibfield  {author} {\bibinfo {author} {\bibfnamefont {R.~F.}\ \bibnamefont
  {Lebed}}, \bibinfo {author} {\bibfnamefont {R.~E.}\ \bibnamefont {Mitchell}},
  \ and\ \bibinfo {author} {\bibfnamefont {E.~S.}\ \bibnamefont {Swanson}},\
  }\href {\doibase 10.1016/j.ppnp.2016.11.003} {\bibfield  {journal} {\bibinfo
  {journal} {Prog. Part. Nucl. Phys.}\ }\textbf {\bibinfo {volume} {93}},\
  \bibinfo {pages} {143} (\bibinfo {year} {2017})},\ \Eprint
  {http://arxiv.org/abs/1610.04528} {arXiv:1610.04528 [hep-ph]} \BibitemShut
  {NoStop}%
\bibitem [{\citenamefont {Briceno}\ \emph {et~al.}(2016)\citenamefont {Briceno}
  \emph {et~al.}}]{Briceno:2015rlt}%
  \BibitemOpen
  \bibfield  {author} {\bibinfo {author} {\bibfnamefont {R.~A.}\ \bibnamefont
  {Briceno}} \emph {et~al.},\ }\href {\doibase 10.1088/1674-1137/40/4/042001}
  {\bibfield  {journal} {\bibinfo  {journal} {Chin. Phys.}\ }\textbf {\bibinfo
  {volume} {C40}},\ \bibinfo {pages} {042001} (\bibinfo {year} {2016})},\
  \Eprint {http://arxiv.org/abs/1511.06779} {arXiv:1511.06779 [hep-ph]}
  \BibitemShut {NoStop}%
\bibitem [{\citenamefont {Dudek}\ \emph {et~al.}(2008)\citenamefont {Dudek},
  \citenamefont {Edwards}, \citenamefont {Mathur},\ and\ \citenamefont
  {Richards}}]{Dudek:2007wv}%
  \BibitemOpen
  \bibfield  {author} {\bibinfo {author} {\bibfnamefont {J.~J.}\ \bibnamefont
  {Dudek}}, \bibinfo {author} {\bibfnamefont {R.~G.}\ \bibnamefont {Edwards}},
  \bibinfo {author} {\bibfnamefont {N.}~\bibnamefont {Mathur}}, \ and\ \bibinfo
  {author} {\bibfnamefont {D.~G.}\ \bibnamefont {Richards}},\ }\href {\doibase
  10.1103/PhysRevD.77.034501} {\bibfield  {journal} {\bibinfo  {journal} {Phys.
  Rev.}\ }\textbf {\bibinfo {volume} {D77}},\ \bibinfo {pages} {034501}
  (\bibinfo {year} {2008})},\ \Eprint {http://arxiv.org/abs/0707.4162}
  {arXiv:0707.4162 [hep-lat]} \BibitemShut {NoStop}%
\bibitem [{\citenamefont {Bali}\ \emph
  {et~al.}(2011{\natexlab{a}})\citenamefont {Bali} \emph
  {et~al.}}]{Bali:2011dc}%
  \BibitemOpen
  \bibfield  {author} {\bibinfo {author} {\bibfnamefont {G.}~\bibnamefont
  {Bali}} \emph {et~al.},\ }\bibfield  {booktitle} {\emph {\bibinfo {booktitle}
  {{Proceedings, 29th International Symposium on Lattice field theory (Lattice
  2011): Squaw Valley, Lake Tahoe, USA, July 10-16, 2011}}},\ }\href@noop {}
  {\bibfield  {journal} {\bibinfo  {journal} {PoS}\ }\textbf {\bibinfo {volume}
  {LATTICE2011}},\ \bibinfo {pages} {135} (\bibinfo {year}
  {2011}{\natexlab{a}})},\ \Eprint {http://arxiv.org/abs/1108.6147}
  {arXiv:1108.6147 [hep-lat]} \BibitemShut {NoStop}%
\bibitem [{\citenamefont {Bali}\ \emph
  {et~al.}(2011{\natexlab{b}})\citenamefont {Bali}, \citenamefont {Collins},\
  and\ \citenamefont {Ehmann}}]{Bali:2011rd}%
  \BibitemOpen
  \bibfield  {author} {\bibinfo {author} {\bibfnamefont {G.~S.}\ \bibnamefont
  {Bali}}, \bibinfo {author} {\bibfnamefont {S.}~\bibnamefont {Collins}}, \
  and\ \bibinfo {author} {\bibfnamefont {C.}~\bibnamefont {Ehmann}},\ }\href
  {\doibase 10.1103/PhysRevD.84.094506} {\bibfield  {journal} {\bibinfo
  {journal} {Phys. Rev.}\ }\textbf {\bibinfo {volume} {D84}},\ \bibinfo {pages}
  {094506} (\bibinfo {year} {2011}{\natexlab{b}})},\ \Eprint
  {http://arxiv.org/abs/1110.2381} {arXiv:1110.2381 [hep-lat]} \BibitemShut
  {NoStop}%
\bibitem [{\citenamefont {Liu}\ \emph {et~al.}(2012)\citenamefont {Liu},
  \citenamefont {Moir}, \citenamefont {Peardon}, \citenamefont {Ryan},
  \citenamefont {Thomas}, \citenamefont {Vilaseca}, \citenamefont {Dudek},
  \citenamefont {Edwards}, \citenamefont {Joo},\ and\ \citenamefont
  {Richards}}]{Liu:2012ze}%
  \BibitemOpen
  \bibfield  {author} {\bibinfo {author} {\bibfnamefont {L.}~\bibnamefont
  {Liu}}, \bibinfo {author} {\bibfnamefont {G.}~\bibnamefont {Moir}}, \bibinfo
  {author} {\bibfnamefont {M.}~\bibnamefont {Peardon}}, \bibinfo {author}
  {\bibfnamefont {S.~M.}\ \bibnamefont {Ryan}}, \bibinfo {author}
  {\bibfnamefont {C.~E.}\ \bibnamefont {Thomas}}, \bibinfo {author}
  {\bibfnamefont {P.}~\bibnamefont {Vilaseca}}, \bibinfo {author}
  {\bibfnamefont {J.~J.}\ \bibnamefont {Dudek}}, \bibinfo {author}
  {\bibfnamefont {R.~G.}\ \bibnamefont {Edwards}}, \bibinfo {author}
  {\bibfnamefont {B.}~\bibnamefont {Joo}}, \ and\ \bibinfo {author}
  {\bibfnamefont {D.~G.}\ \bibnamefont {Richards}} (\bibinfo {collaboration}
  {Hadron Spectrum}),\ }\href {\doibase 10.1007/JHEP07(2012)126} {\bibfield
  {journal} {\bibinfo  {journal} {JHEP}\ }\textbf {\bibinfo {volume} {07}},\
  \bibinfo {pages} {126} (\bibinfo {year} {2012})},\ \Eprint
  {http://arxiv.org/abs/1204.5425} {arXiv:1204.5425 [hep-ph]} \BibitemShut
  {NoStop}%
\bibitem [{\citenamefont {Cheung}\ \emph {et~al.}(2016)\citenamefont {Cheung},
  \citenamefont {O'Hara}, \citenamefont {Moir}, \citenamefont {Peardon},
  \citenamefont {Ryan}, \citenamefont {Thomas},\ and\ \citenamefont
  {Tims}}]{Cheung:2016bym}%
  \BibitemOpen
  \bibfield  {author} {\bibinfo {author} {\bibfnamefont {G.~K.~C.}\
  \bibnamefont {Cheung}}, \bibinfo {author} {\bibfnamefont {C.}~\bibnamefont
  {O'Hara}}, \bibinfo {author} {\bibfnamefont {G.}~\bibnamefont {Moir}},
  \bibinfo {author} {\bibfnamefont {M.}~\bibnamefont {Peardon}}, \bibinfo
  {author} {\bibfnamefont {S.~M.}\ \bibnamefont {Ryan}}, \bibinfo {author}
  {\bibfnamefont {C.~E.}\ \bibnamefont {Thomas}}, \ and\ \bibinfo {author}
  {\bibfnamefont {D.}~\bibnamefont {Tims}} (\bibinfo {collaboration} {Hadron
  Spectrum}),\ }\href {\doibase 10.1007/JHEP12(2016)089} {\bibfield  {journal}
  {\bibinfo  {journal} {JHEP}\ }\textbf {\bibinfo {volume} {12}},\ \bibinfo
  {pages} {089} (\bibinfo {year} {2016})},\ \Eprint
  {http://arxiv.org/abs/1610.01073} {arXiv:1610.01073 [hep-lat]} \BibitemShut
  {NoStop}%
\bibitem [{\citenamefont {Juge}\ \emph {et~al.}(1998)\citenamefont {Juge},
  \citenamefont {Kuti},\ and\ \citenamefont {Morningstar}}]{Juge:1997nc}%
  \BibitemOpen
  \bibfield  {author} {\bibinfo {author} {\bibfnamefont {K.~J.}\ \bibnamefont
  {Juge}}, \bibinfo {author} {\bibfnamefont {J.}~\bibnamefont {Kuti}}, \ and\
  \bibinfo {author} {\bibfnamefont {C.~J.}\ \bibnamefont {Morningstar}},\
  }\bibfield  {booktitle} {\emph {\bibinfo {booktitle} {{Contents of LAT97
  proceedings}}},\ }\href {\doibase 10.1016/S0920-5632(97)00759-7} {\bibfield
  {journal} {\bibinfo  {journal} {Nucl. Phys. Proc. Suppl.}\ }\textbf {\bibinfo
  {volume} {63}},\ \bibinfo {pages} {326} (\bibinfo {year} {1998})},\ \Eprint
  {http://arxiv.org/abs/hep-lat/9709131} {arXiv:hep-lat/9709131 [hep-lat]}
  \BibitemShut {NoStop}%
\bibitem [{\citenamefont {Bali}\ \emph {et~al.}(2000)\citenamefont {Bali},
  \citenamefont {Bolder}, \citenamefont {Eicker}, \citenamefont {Lippert},
  \citenamefont {Orth}, \citenamefont {Ueberholz}, \citenamefont {Schilling},\
  and\ \citenamefont {Struckmann}}]{Bali:2000vr}%
  \BibitemOpen
  \bibfield  {author} {\bibinfo {author} {\bibfnamefont {G.~S.}\ \bibnamefont
  {Bali}}, \bibinfo {author} {\bibfnamefont {B.}~\bibnamefont {Bolder}},
  \bibinfo {author} {\bibfnamefont {N.}~\bibnamefont {Eicker}}, \bibinfo
  {author} {\bibfnamefont {T.}~\bibnamefont {Lippert}}, \bibinfo {author}
  {\bibfnamefont {B.}~\bibnamefont {Orth}}, \bibinfo {author} {\bibfnamefont
  {P.}~\bibnamefont {Ueberholz}}, \bibinfo {author} {\bibfnamefont
  {K.}~\bibnamefont {Schilling}}, \ and\ \bibinfo {author} {\bibfnamefont
  {T.}~\bibnamefont {Struckmann}} (\bibinfo {collaboration} {TXL, T(X)L}),\
  }\href {\doibase 10.1103/PhysRevD.62.054503} {\bibfield  {journal} {\bibinfo
  {journal} {Phys. Rev.}\ }\textbf {\bibinfo {volume} {D62}},\ \bibinfo {pages}
  {054503} (\bibinfo {year} {2000})},\ \Eprint
  {http://arxiv.org/abs/hep-lat/0003012} {arXiv:hep-lat/0003012 [hep-lat]}
  \BibitemShut {NoStop}%
\bibitem [{\citenamefont {Juge}\ \emph {et~al.}(2003)\citenamefont {Juge},
  \citenamefont {Kuti},\ and\ \citenamefont {Morningstar}}]{Juge:2002br}%
  \BibitemOpen
  \bibfield  {author} {\bibinfo {author} {\bibfnamefont {K.~J.}\ \bibnamefont
  {Juge}}, \bibinfo {author} {\bibfnamefont {J.}~\bibnamefont {Kuti}}, \ and\
  \bibinfo {author} {\bibfnamefont {C.}~\bibnamefont {Morningstar}},\ }\href
  {\doibase 10.1103/PhysRevLett.90.161601} {\bibfield  {journal} {\bibinfo
  {journal} {Phys. Rev. Lett.}\ }\textbf {\bibinfo {volume} {90}},\ \bibinfo
  {pages} {161601} (\bibinfo {year} {2003})},\ \Eprint
  {http://arxiv.org/abs/hep-lat/0207004} {arXiv:hep-lat/0207004 [hep-lat]}
  \BibitemShut {NoStop}%
\bibitem [{\citenamefont {Bali}\ and\ \citenamefont
  {Pineda}(2004)}]{Bali:2003jq}%
  \BibitemOpen
  \bibfield  {author} {\bibinfo {author} {\bibfnamefont {G.~S.}\ \bibnamefont
  {Bali}}\ and\ \bibinfo {author} {\bibfnamefont {A.}~\bibnamefont {Pineda}},\
  }\href {\doibase 10.1103/PhysRevD.69.094001} {\bibfield  {journal} {\bibinfo
  {journal} {Phys. Rev.}\ }\textbf {\bibinfo {volume} {D69}},\ \bibinfo {pages}
  {094001} (\bibinfo {year} {2004})},\ \Eprint
  {http://arxiv.org/abs/hep-ph/0310130} {arXiv:hep-ph/0310130 [hep-ph]}
  \BibitemShut {NoStop}%
\bibitem [{\citenamefont {Brambilla}\ \emph
  {et~al.}(2005{\natexlab{b}})\citenamefont {Brambilla}, \citenamefont
  {Pineda}, \citenamefont {Soto},\ and\ \citenamefont
  {Vairo}}]{Brambilla:2004jw}%
  \BibitemOpen
  \bibfield  {author} {\bibinfo {author} {\bibfnamefont {N.}~\bibnamefont
  {Brambilla}}, \bibinfo {author} {\bibfnamefont {A.}~\bibnamefont {Pineda}},
  \bibinfo {author} {\bibfnamefont {J.}~\bibnamefont {Soto}}, \ and\ \bibinfo
  {author} {\bibfnamefont {A.}~\bibnamefont {Vairo}},\ }\href {\doibase
  10.1103/RevModPhys.77.1423} {\bibfield  {journal} {\bibinfo  {journal} {Rev.
  Mod. Phys.}\ }\textbf {\bibinfo {volume} {77}},\ \bibinfo {pages} {1423}
  (\bibinfo {year} {2005}{\natexlab{b}})},\ \Eprint
  {http://arxiv.org/abs/hep-ph/0410047} {arXiv:hep-ph/0410047 [hep-ph]}
  \BibitemShut {NoStop}%
\bibitem [{\citenamefont {Griffiths}\ \emph {et~al.}(1983)\citenamefont
  {Griffiths}, \citenamefont {Michael},\ and\ \citenamefont
  {Rakow}}]{Griffiths:1983ah}%
  \BibitemOpen
  \bibfield  {author} {\bibinfo {author} {\bibfnamefont {L.~A.}\ \bibnamefont
  {Griffiths}}, \bibinfo {author} {\bibfnamefont {C.}~\bibnamefont {Michael}},
  \ and\ \bibinfo {author} {\bibfnamefont {P.~E.~L.}\ \bibnamefont {Rakow}},\
  }\href {\doibase 10.1016/0370-2693(83)90680-9} {\bibfield  {journal}
  {\bibinfo  {journal} {Phys. Lett.}\ }\textbf {\bibinfo {volume} {129B}},\
  \bibinfo {pages} {351} (\bibinfo {year} {1983})}\BibitemShut {NoStop}%
\bibitem [{\citenamefont {Juge}\ \emph {et~al.}(1999)\citenamefont {Juge},
  \citenamefont {Kuti},\ and\ \citenamefont {Morningstar}}]{Juge:1999ie}%
  \BibitemOpen
  \bibfield  {author} {\bibinfo {author} {\bibfnamefont {K.~J.}\ \bibnamefont
  {Juge}}, \bibinfo {author} {\bibfnamefont {J.}~\bibnamefont {Kuti}}, \ and\
  \bibinfo {author} {\bibfnamefont {C.~J.}\ \bibnamefont {Morningstar}},\
  }\href {\doibase 10.1103/PhysRevLett.82.4400} {\bibfield  {journal} {\bibinfo
   {journal} {Phys. Rev. Lett.}\ }\textbf {\bibinfo {volume} {82}},\ \bibinfo
  {pages} {4400} (\bibinfo {year} {1999})},\ \Eprint
  {http://arxiv.org/abs/hep-ph/9902336} {arXiv:hep-ph/9902336 [hep-ph]}
  \BibitemShut {NoStop}%
\bibitem [{\citenamefont {Braaten}\ \emph
  {et~al.}(2014{\natexlab{a}})\citenamefont {Braaten}, \citenamefont
  {Langmack},\ and\ \citenamefont {Smith}}]{Braaten:2014qka}%
  \BibitemOpen
  \bibfield  {author} {\bibinfo {author} {\bibfnamefont {E.}~\bibnamefont
  {Braaten}}, \bibinfo {author} {\bibfnamefont {C.}~\bibnamefont {Langmack}}, \
  and\ \bibinfo {author} {\bibfnamefont {D.~H.}\ \bibnamefont {Smith}},\ }\href
  {\doibase 10.1103/PhysRevD.90.014044} {\bibfield  {journal} {\bibinfo
  {journal} {Phys. Rev.}\ }\textbf {\bibinfo {volume} {D90}},\ \bibinfo {pages}
  {014044} (\bibinfo {year} {2014}{\natexlab{a}})},\ \Eprint
  {http://arxiv.org/abs/1402.0438} {arXiv:1402.0438 [hep-ph]} \BibitemShut
  {NoStop}%
\bibitem [{\citenamefont {Braaten}\ \emph
  {et~al.}(2014{\natexlab{b}})\citenamefont {Braaten}, \citenamefont
  {Langmack},\ and\ \citenamefont {Smith}}]{Braaten:2014ita}%
  \BibitemOpen
  \bibfield  {author} {\bibinfo {author} {\bibfnamefont {E.}~\bibnamefont
  {Braaten}}, \bibinfo {author} {\bibfnamefont {C.}~\bibnamefont {Langmack}}, \
  and\ \bibinfo {author} {\bibfnamefont {D.~H.}\ \bibnamefont {Smith}},\ }\href
  {\doibase 10.1103/PhysRevLett.112.222001} {\bibfield  {journal} {\bibinfo
  {journal} {Phys. Rev. Lett.}\ }\textbf {\bibinfo {volume} {112}},\ \bibinfo
  {pages} {222001} (\bibinfo {year} {2014}{\natexlab{b}})},\ \Eprint
  {http://arxiv.org/abs/1401.7351} {arXiv:1401.7351 [hep-ph]} \BibitemShut
  {NoStop}%
\bibitem [{\citenamefont {Berwein}\ \emph {et~al.}(2015)\citenamefont
  {Berwein}, \citenamefont {Brambilla}, \citenamefont {Tarr\'us~Castell\`a},\
  and\ \citenamefont {Vairo}}]{Berwein:2015vca}%
  \BibitemOpen
  \bibfield  {author} {\bibinfo {author} {\bibfnamefont {M.}~\bibnamefont
  {Berwein}}, \bibinfo {author} {\bibfnamefont {N.}~\bibnamefont {Brambilla}},
  \bibinfo {author} {\bibfnamefont {J.}~\bibnamefont {Tarr\'us~Castell\`a}}, \
  and\ \bibinfo {author} {\bibfnamefont {A.}~\bibnamefont {Vairo}},\ }\href
  {\doibase 10.1103/PhysRevD.92.114019} {\bibfield  {journal} {\bibinfo
  {journal} {Phys. Rev.}\ }\textbf {\bibinfo {volume} {D92}},\ \bibinfo {pages}
  {114019} (\bibinfo {year} {2015})},\ \Eprint
  {http://arxiv.org/abs/1510.04299} {arXiv:1510.04299 [hep-ph]} \BibitemShut
  {NoStop}%
\bibitem [{\citenamefont {Oncala}\ and\ \citenamefont
  {Soto}(2017)}]{Oncala:2017hop}%
  \BibitemOpen
  \bibfield  {author} {\bibinfo {author} {\bibfnamefont {R.}~\bibnamefont
  {Oncala}}\ and\ \bibinfo {author} {\bibfnamefont {J.}~\bibnamefont {Soto}},\
  }\href {\doibase 10.1103/PhysRevD.96.014004} {\bibfield  {journal} {\bibinfo
  {journal} {Phys. Rev.}\ }\textbf {\bibinfo {volume} {D96}},\ \bibinfo {pages}
  {014004} (\bibinfo {year} {2017})},\ \Eprint
  {http://arxiv.org/abs/1702.03900} {arXiv:1702.03900 [hep-ph]} \BibitemShut
  {NoStop}%
\bibitem [{\citenamefont {Brambilla}\ \emph {et~al.}(2018)\citenamefont
  {Brambilla}, \citenamefont {Krein}, \citenamefont {Tarr{\'u}s~Castell{\`a}},\
  and\ \citenamefont {Vairo}}]{Brambilla:2017uyf}%
  \BibitemOpen
  \bibfield  {author} {\bibinfo {author} {\bibfnamefont {N.}~\bibnamefont
  {Brambilla}}, \bibinfo {author} {\bibfnamefont {G.}~\bibnamefont {Krein}},
  \bibinfo {author} {\bibfnamefont {J.}~\bibnamefont
  {Tarr{\'u}s~Castell{\`a}}}, \ and\ \bibinfo {author} {\bibfnamefont
  {A.}~\bibnamefont {Vairo}},\ }\href {\doibase 10.1103/PhysRevD.97.016016}
  {\bibfield  {journal} {\bibinfo  {journal} {Phys. Rev.}\ }\textbf {\bibinfo
  {volume} {D97}},\ \bibinfo {pages} {016016} (\bibinfo {year} {2018})},\
  \Eprint {http://arxiv.org/abs/1707.09647} {arXiv:1707.09647 [hep-ph]}
  \BibitemShut {NoStop}%
\bibitem [{\citenamefont {Soto}(2018)}]{Soto:2017one}%
  \BibitemOpen
  \bibfield  {author} {\bibinfo {author} {\bibfnamefont {J.}~\bibnamefont
  {Soto}},\ }\bibfield  {booktitle} {\emph {\bibinfo {booktitle} {{Proceedings,
  20th High-Energy Physics International Conference in Quantum Chromodynamics
  (QCD 17): Montpellier, France, July 3-7, 2017}}},\ }\href {\doibase
  10.1016/j.nuclphysbps.2018.03.020} {\bibfield  {journal} {\bibinfo  {journal}
  {Nucl. Part. Phys. Proc.}\ }\textbf {\bibinfo {volume} {294-296}},\ \bibinfo
  {pages} {87} (\bibinfo {year} {2018})},\ \Eprint
  {http://arxiv.org/abs/1709.08038} {arXiv:1709.08038 [hep-ph]} \BibitemShut
  {NoStop}%
\bibitem [{\citenamefont {Foster}\ and\ \citenamefont
  {Michael}(1999)}]{Foster:1998wu}%
  \BibitemOpen
  \bibfield  {author} {\bibinfo {author} {\bibfnamefont {M.}~\bibnamefont
  {Foster}}\ and\ \bibinfo {author} {\bibfnamefont {C.}~\bibnamefont {Michael}}
  (\bibinfo {collaboration} {UKQCD}),\ }\href {\doibase
  10.1103/PhysRevD.59.094509} {\bibfield  {journal} {\bibinfo  {journal} {Phys.
  Rev.}\ }\textbf {\bibinfo {volume} {D59}},\ \bibinfo {pages} {094509}
  (\bibinfo {year} {1999})},\ \Eprint {http://arxiv.org/abs/hep-lat/9811010}
  {arXiv:hep-lat/9811010 [hep-lat]} \BibitemShut {NoStop}%
\bibitem [{\citenamefont {Brambilla}\ \emph
  {et~al.}(2000{\natexlab{a}})\citenamefont {Brambilla}, \citenamefont
  {Pineda}, \citenamefont {Soto},\ and\ \citenamefont
  {Vairo}}]{Brambilla:1999xf}%
  \BibitemOpen
  \bibfield  {author} {\bibinfo {author} {\bibfnamefont {N.}~\bibnamefont
  {Brambilla}}, \bibinfo {author} {\bibfnamefont {A.}~\bibnamefont {Pineda}},
  \bibinfo {author} {\bibfnamefont {J.}~\bibnamefont {Soto}}, \ and\ \bibinfo
  {author} {\bibfnamefont {A.}~\bibnamefont {Vairo}},\ }\href {\doibase
  10.1016/S0550-3213(99)00693-8} {\bibfield  {journal} {\bibinfo  {journal}
  {Nucl. Phys.}\ }\textbf {\bibinfo {volume} {B566}},\ \bibinfo {pages} {275}
  (\bibinfo {year} {2000}{\natexlab{a}})},\ \Eprint
  {http://arxiv.org/abs/hep-ph/9907240} {arXiv:hep-ph/9907240 [hep-ph]}
  \BibitemShut {NoStop}%
\bibitem [{\citenamefont {Pineda}(1997)}]{Pineda:1996nw}%
  \BibitemOpen
  \bibfield  {author} {\bibinfo {author} {\bibfnamefont {A.}~\bibnamefont
  {Pineda}},\ }\href {\doibase 10.1103/PhysRevD.59.099901,
  10.1103/PhysRevD.55.407} {\bibfield  {journal} {\bibinfo  {journal} {Phys.
  Rev.}\ }\textbf {\bibinfo {volume} {D55}},\ \bibinfo {pages} {407} (\bibinfo
  {year} {1997})},\ \bibinfo {note} {[Erratum: Phys. Rev.D59,099901(1999)]},\
  \Eprint {http://arxiv.org/abs/hep-ph/9605424} {arXiv:hep-ph/9605424 [hep-ph]}
  \BibitemShut {NoStop}%
\bibitem [{\citenamefont {Brambilla}\ \emph {et~al.}(2003)\citenamefont
  {Brambilla}, \citenamefont {Gromes},\ and\ \citenamefont
  {Vairo}}]{Brambilla:2003nt}%
  \BibitemOpen
  \bibfield  {author} {\bibinfo {author} {\bibfnamefont {N.}~\bibnamefont
  {Brambilla}}, \bibinfo {author} {\bibfnamefont {D.}~\bibnamefont {Gromes}}, \
  and\ \bibinfo {author} {\bibfnamefont {A.}~\bibnamefont {Vairo}},\ }\href
  {\doibase 10.1016/j.physletb.2003.09.100} {\bibfield  {journal} {\bibinfo
  {journal} {Phys. Lett.}\ }\textbf {\bibinfo {volume} {B576}},\ \bibinfo
  {pages} {314} (\bibinfo {year} {2003})},\ \Eprint
  {http://arxiv.org/abs/hep-ph/0306107} {arXiv:hep-ph/0306107 [hep-ph]}
  \BibitemShut {NoStop}%
\bibitem [{\citenamefont {Brambilla}\ \emph
  {et~al.}(2000{\natexlab{b}})\citenamefont {Brambilla}, \citenamefont
  {Pineda}, \citenamefont {Soto},\ and\ \citenamefont
  {Vairo}}]{Brambilla:2000gk}%
  \BibitemOpen
  \bibfield  {author} {\bibinfo {author} {\bibfnamefont {N.}~\bibnamefont
  {Brambilla}}, \bibinfo {author} {\bibfnamefont {A.}~\bibnamefont {Pineda}},
  \bibinfo {author} {\bibfnamefont {J.}~\bibnamefont {Soto}}, \ and\ \bibinfo
  {author} {\bibfnamefont {A.}~\bibnamefont {Vairo}},\ }\href {\doibase
  10.1103/PhysRevD.63.014023} {\bibfield  {journal} {\bibinfo  {journal} {Phys.
  Rev.}\ }\textbf {\bibinfo {volume} {D63}},\ \bibinfo {pages} {014023}
  (\bibinfo {year} {2000}{\natexlab{b}})},\ \Eprint
  {http://arxiv.org/abs/hep-ph/0002250} {arXiv:hep-ph/0002250 [hep-ph]}
  \BibitemShut {NoStop}%
\bibitem [{\citenamefont {Eichten}\ and\ \citenamefont
  {Hill}(1990)}]{Eichten:1990vp}%
  \BibitemOpen
  \bibfield  {author} {\bibinfo {author} {\bibfnamefont {E.}~\bibnamefont
  {Eichten}}\ and\ \bibinfo {author} {\bibfnamefont {B.~R.}\ \bibnamefont
  {Hill}},\ }\href {\doibase 10.1016/0370-2693(90)91408-4} {\bibfield
  {journal} {\bibinfo  {journal} {Phys. Lett.}\ }\textbf {\bibinfo {volume}
  {B243}},\ \bibinfo {pages} {427} (\bibinfo {year} {1990})}\BibitemShut
  {NoStop}%
\bibitem [{\citenamefont {Landau}\ and\ \citenamefont
  {Lifshitz}(1977)}]{LandauLifshitz}%
  \BibitemOpen
  \bibfield  {author} {\bibinfo {author} {\bibfnamefont {L.~D.}\ \bibnamefont
  {Landau}}\ and\ \bibinfo {author} {\bibfnamefont {E.~M.}\ \bibnamefont
  {Lifshitz}},\ }\href@noop {} {\emph {\bibinfo {title} {{Quantum Mechanics:
  Non-Relativistic Theory}}}},\ \bibinfo {edition} {3rd}\ ed.,\ \bibinfo
  {series} {Course of Theoretical Physics}, Vol.~\bibinfo {volume} {3}\
  (\bibinfo  {publisher} {Pergamon Press},\ \bibinfo {year} {1977})\BibitemShut
  {NoStop}%
\bibitem [{\citenamefont {Pineda}(2001)}]{Pineda:2001zq}%
  \BibitemOpen
  \bibfield  {author} {\bibinfo {author} {\bibfnamefont {A.}~\bibnamefont
  {Pineda}},\ }\href {\doibase 10.1088/1126-6708/2001/06/022} {\bibfield
  {journal} {\bibinfo  {journal} {JHEP}\ }\textbf {\bibinfo {volume} {06}},\
  \bibinfo {pages} {022} (\bibinfo {year} {2001})},\ \Eprint
  {http://arxiv.org/abs/hep-ph/0105008} {arXiv:hep-ph/0105008 [hep-ph]}
  \BibitemShut {NoStop}%
\end{thebibliography}%

\end{document}